\documentclass[reprint,amsmath,amssymb,aps,pra]{revtex4-2}

\usepackage[utf8]{inputenc}
\usepackage[T1]{fontenc}
\usepackage{graphicx}
\usepackage{dcolumn}
\usepackage{bm}
\usepackage{color}
\usepackage[colorlinks=true,linkcolor=blue,urlcolor=blue,citecolor=blue,pdfusetitle]{hyperref}
\usepackage{upgreek,textgreek}
\usepackage{physics}
\usepackage{orcidlink}
\usepackage{booktabs}

\newcommand{\Hop}{\hat{H}}
\newcommand{\aop}{\hat{a}}
\newcommand{\aopd}{\hat{a}^\dag}
\newcommand{\nop}{\hat{n}}

\newcommand{\UAB}{U_{AB}}
\newcommand{\eb}{\epsilon_{b}}

\begin{document}

\title{Few-body bound states of bosonic mixtures in two-dimensional optical lattices}

\author{Matias Volante-Abovich~\orcidlink{0009-0001-8713-3062}}
\author{Felipe Isaule~\orcidlink{0000-0003-1810-0707}}
\author{Luis Morales-Molina}

\affiliation{Instituto de Física, Pontificia
Universidad Católica de Chile,
Avenida Vicuña Mackenna 4860,
Santiago, Chile.}

\date{\today}

\begin{abstract}

We study the formation of bound states in a binary mixture of a few bosons in small square optical lattices. Using the exact diagonalization method, we find that bound clusters of all available bosons can form. We provide a comprehensive numerical examination of these bound states for a wide range of repulsive intraspecies and attractive interspecies interactions. In contrast to binary mixtures in one-dimensional chains, we reveal that the binding energy of the clusters shows a non-monotonic dependence on the interspecies interaction strengths for small tunneling rates, developing a local minimum for intermediate attractive interactions. 
The findings of this work highlight the difference between the binding mechanisms of binary bosonic mixtures in one- and higher-dimensional lattices.

\end{abstract}

\maketitle

\section{Introduction}

Mixtures of ultracold bosonic atoms have received increased attention during the past two decades~\cite{sowinski_one-dimensional_2019,baroni_quantum_2024}. The interplay between the different atomic species can lead to a plethora of rich physics, such as the superfluid drag~\cite{fil_nondissipative_2005,nespolo_andreevbashkin_2017},  Josephson oscillations~\cite{julia-diaz_josephson_2009,mele-messeguer_weakly_2011}, and coherent currents~\cite{morales-molina_harnessing_2020,spehner_persistent_2021}. Within these, of particular interest is the study of quantum bound configurations in attractive bosonic mixtures, ranging from few-body clusters~\cite{guijarro_one-dimensional_2018,guijarro_few-body_2020,liu_three-body_2021} to many-body droplets and liquids~\cite{petrov_quantum_2015,petrov_ultradilute_2016}.
Droplets are a seminal example of the importance of quantum fluctuations, as they are a fully beyond mean-field (MF) effect. Indeed, while MF theory predicts the collapse of binary bosonic gases for strong attractive interactions, quantum fluctuations prevent such collapse, forming a stable quantum droplet instead~\cite{petrov_quantum_2015}. Since their predictions, droplets have been successfully realized in a series of landmark experiments with ultracold bosonic mixtures~\cite{cabrera_quantum_2018,semeghini_self-bound_2018,cheiney_bright_2018,ferioli_collisions_2019,derrico_observation_2019}. In turn, these developments have motivated extensive theoretical studies of quantum droplets and liquids~\cite{hu_consistent_2020,hu_microscopic_2020-1,ota_beyond_2020,tylutki_collective_2020,guebli_quantum_2021,xiong_effective_2022,pan_quantum_2022,boudjemaa_quantum_2023,spada_quantum_2024}. Furthermore, similar physics can also be explored with dipolar gases~\cite{lahaye_physics_2009,baranov_condensed_2012,trefzger_ultracold_2011,chomaz_dipolar_2023}, which have led to the observation of dipolar droplets~\cite{schmitt_self-bound_2016,ferrier-barbut_observation_2016,chomaz_quantum-fluctuation-driven_2016} and multiple theoretical studies of few-body bound states~\cite{klawunn_two-dimensional_2010,wunsch_few-body_2011,volosniev_few-body_2012,volosniev_bound_2013}.

One useful platform for studying clustering phenomena is optical lattices~\cite{bloch_ultracold_2005,lewenstein_ultracold_2012,gross_quantum_2017}. Theoretically, ultracold atoms trapped in optical lattices can be described with high accuracy in the tight-binding approximation by Hubbard-based models~\cite{jaksch_cold_2005}. In such models, both the tunneling parameters and on-site interactions can be controlled in ultracold atom experiments~\cite{chin_feshbach_2010}. In this direction, quantum droplets, pair superfluids, and few-body bound states have already been predicted to form in binary bosonic mixtures confined in one-dimensional lattices~\cite{morera_quantum_2020,morera_universal_2021,machida_quantum_2022,boudjemaa_discrete_2023,valles-muns_quantum_2024}.  Closely related, the formation of bound trimers in three-component fermionic systems in optical lattices has also been investigated~\cite{rapp_color_2007,rapp_trionic_2008,klingschat_exact_2010,backes_monte_2012,pohlmann_trion_2013,xu_trion_2023}, showing that the dimensionality can play an important role. Thus, a further examination of bosonic mixtures in two- and three-dimensional lattices could elucidate the dependence of clustering on the lattice geometry.

In this work, we examine the formation of bound states in binary mixtures of a few bosons confined in small two-dimensional square lattices. We consider three to six particles and study the system numerically using the exact diagonalization (ED) method~\cite{zhang_exact_2010,raventos_cold_2017}, focusing mainly on ground-state properties.
In the considered small lattices, we find that bound tetramers and hexamers are formed for attractive interspecies interactions. Importantly, for small tunneling rates, the binding energy shows a non-monotonic dependence on the interspecies attraction, thus developing a local minimum due to quantum effects. The latter behavior is similar to what has been found in few-body binary mixtures in the continuum~\cite{guijarro_few-body_2020}.  However, such non-monotonic dependence is not found in one-dimensional chains, showing that mixtures in lattices with a larger coordination number behave very differently from the one-dimensional mixtures studied in the past~\cite{morera_quantum_2020,morera_universal_2021}. We present a comprehensive study of the bound states by examining the binding energies, interparticle distances, and entanglement properties across a wide range of interaction strengths.    

This work is organized as follows. Sec.~\ref{sec:model} presents the physical model and the employed numerical approach. Sec.~\ref{sec:results} presents the main results of this work, corresponding to ground-state properties of balanced mixtures with two and three bosons of each atomic species. Binding energies, average distances, and von Neumann entropies are examined in this section. Then, Sec.~\ref{sec:concl} presents the conclusions of this work, as well as an outlook for future directions. Additional results are also provided in the Appendix.

\section{Model}
\label{sec:model}

\subsection{Hamiltonian}

\begin{figure}
    \centering
    \includegraphics[width=\columnwidth]{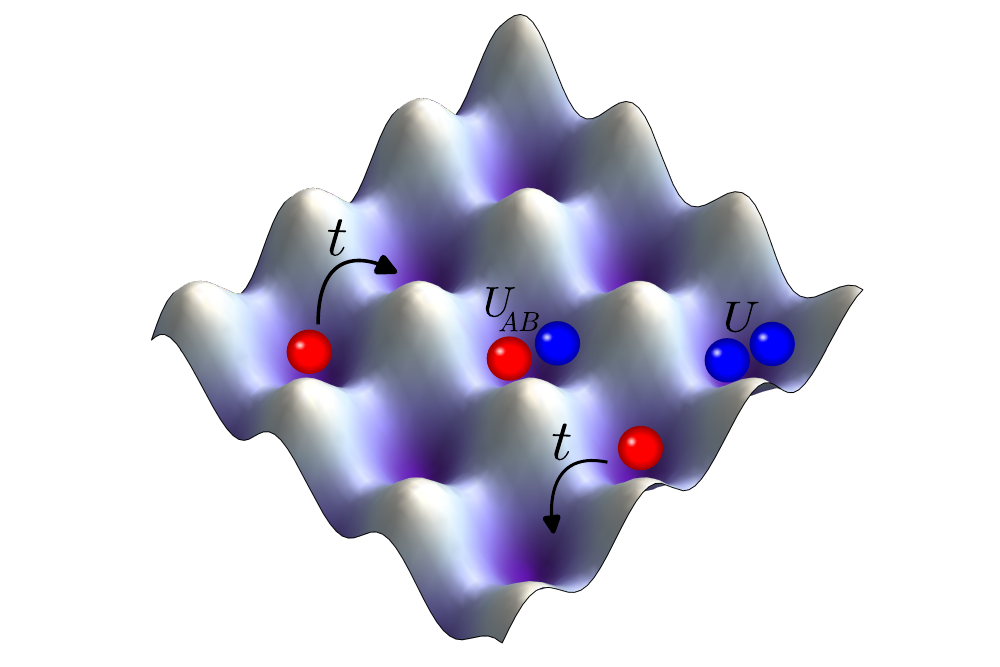}
    \caption{Illustration of the system under consideration, where a few bosons of species $A$ (red circles) and $B$ (blue circles) are trapped in a square optical lattice.} 
    \label{sec:model;sub:H;fig:illustration}
\end{figure}

We consider a tight square two-dimensional optical lattice with $M=M_x\times M_y$ sites. The lattice is loaded with two species of bosons, $A$ and $B$, which interact through short-range on-site potentials. We illustrate the system under consideration in Fig.~\ref{sec:model;sub:H;fig:illustration}. The system is modelled with a two-component Bose-Hubbard Hamiltonian~\cite{jaksch_cold_1998}
\begin{align}
    \Hop =& -t\sum_{\sigma=A,B}\sum_{\langle i,j \rangle}\left(\aopd_{i,\sigma}\aop_{j,\sigma}+\textrm{h.c.}\right)\nonumber\\
    &+\frac{U}{2}\sum_{\sigma=A,B}\sum_i \nop_{i,\sigma}\left(\nop_{i,\sigma}-1\right)\nonumber\\
    &+\UAB\sum_i  \nop_{i,A} \nop_{i,B},
    \label{sec:model;sub:H;eq:H}
\end{align}
where $\aop_{i,\sigma}$ ( $\aopd_{i,\sigma}$) annihilates (creates) a boson of species $\sigma$ at site $i$, and $\nop_{i,\sigma}=\aopd_{i,\sigma}\aop_{i,\sigma}$ is the number operator. The first term in the Hamiltonian describes the tunneling to nearest-neighbor sites, with $\langle i,j \rangle$ referring to neighboring sites in the lattice. The second and third terms describe the intraspecies and interspecies on-site interaction, respectively. Note that we consider equal tunneling coupling $t$ and intraspecies interaction strength $U$ for both species. 

In this work, we consider repulsive intraspecies interactions $U>0$, while we consider attractive interspecies ones $\UAB<0$. We mostly focus on the region $|\UAB|/U\leq 1$ to study the formation of bound states induced by quantum fluctuations. The region $|\UAB|/U> 1$ is trivial, as the system simply collapses to one site due to the strong attraction between bosons.

\subsection{Exact diagonalization}

To study our model, we employ the exact diagonalization (ED) method for a fixed number of particles of each species $N_A$ and $N_B$~\cite{zhang_exact_2010,raventos_cold_2017}. While ED restricts our study to small lattices with a few particles, it enables us to study the system with high accuracy and easily access all the relevant physical properties. We refer to Refs.~\cite{klingschat_exact_2010,pohlmann_trion_2013} for related ED studies of fermionic mixtures in two-dimensional lattices.

The main text of this work considers balanced mixtures with $N_A=N_B=2$ (AABB) and $N_A=N_B=3$ (AAABBB) bosons and lattices with up to $9\times 9$ and $5\times 5$ sites, respectively. We consider periodic boundary conditions to avoid boundary effects. While two-dimensional periodic lattices are unrealistic, they are able to capture the properties of interest, as done in the previously mentioned ED studies of fermionic mixtures~\cite{klingschat_exact_2010,pohlmann_trion_2013} and, more recently, of Fermi polarons~\cite{amelio_polaron_2024}. Nevertheless, extrapolations to large lattices should be taken with care.

In our two-component system, the Hilbert space $\mathcal{H}$ is given by the tensor product of the space of each species $\mathcal{H}=\mathcal{H}_A\otimes\mathcal{H}_B$.  To construct the Hamiltonian matrices and perform the diagonalizations, we use the standard basis in which each Fock state $|\alpha\rangle \in \mathcal{H}_A$ and $|\beta\rangle \in \mathcal{H}_B$ is labeled by the occupations of the different sites within the lattice~\cite{zhang_exact_2010}. Therefore, a state $|\nu\rangle$ in $\mathcal{H}$ is given by 
\begin{equation}
    |\nu\rangle=|\alpha\rangle |\beta\rangle=|n^{(\nu)}_{1,A},...,n^{(\nu)}_{M,A}\rangle |n^{(\nu)}
    _{1,B},...,n^{(\nu)}_{M,B}\rangle,
\end{equation}
where $n^{(\nu)}_{i,\sigma}$ represents the number of particles of species $\sigma$ in site $i$ for state $|\nu\rangle$, and thus $N_\sigma=\sum_i n^{(\nu)}_{i,\sigma}$. A wavefunction then reads
\begin{equation}
    |\Psi\rangle = \sum^{\mathcal{D}_A}_{\alpha=1}\sum^{\mathcal{D}_B}_{\beta=1} c_{\alpha,\beta} |\alpha\rangle |\beta\rangle=\sum^{\mathcal{D}}_{\nu=1} c_{\nu} |\nu\rangle,
\end{equation}
where $\mathcal{D}_\sigma=\binom{N_\sigma+M-1}{N_\sigma}$ is the number of Fock states for each species~\cite{raventos_cold_2017}. Therefore, $\mathcal{D}=\mathcal{D}_A\mathcal{D}_B$ is the size of the full Hilbert space. The eigenenergies and eigenvectors, with their coefficients $c_{\nu}$, are obtained from standard numerical diagonalization for large sparse matrices~\cite{lehoucq_arpack_1998}.

\subsection{Experimental realization}

\begin{figure*}[t!]
    \centering
    \includegraphics[width=\textwidth]{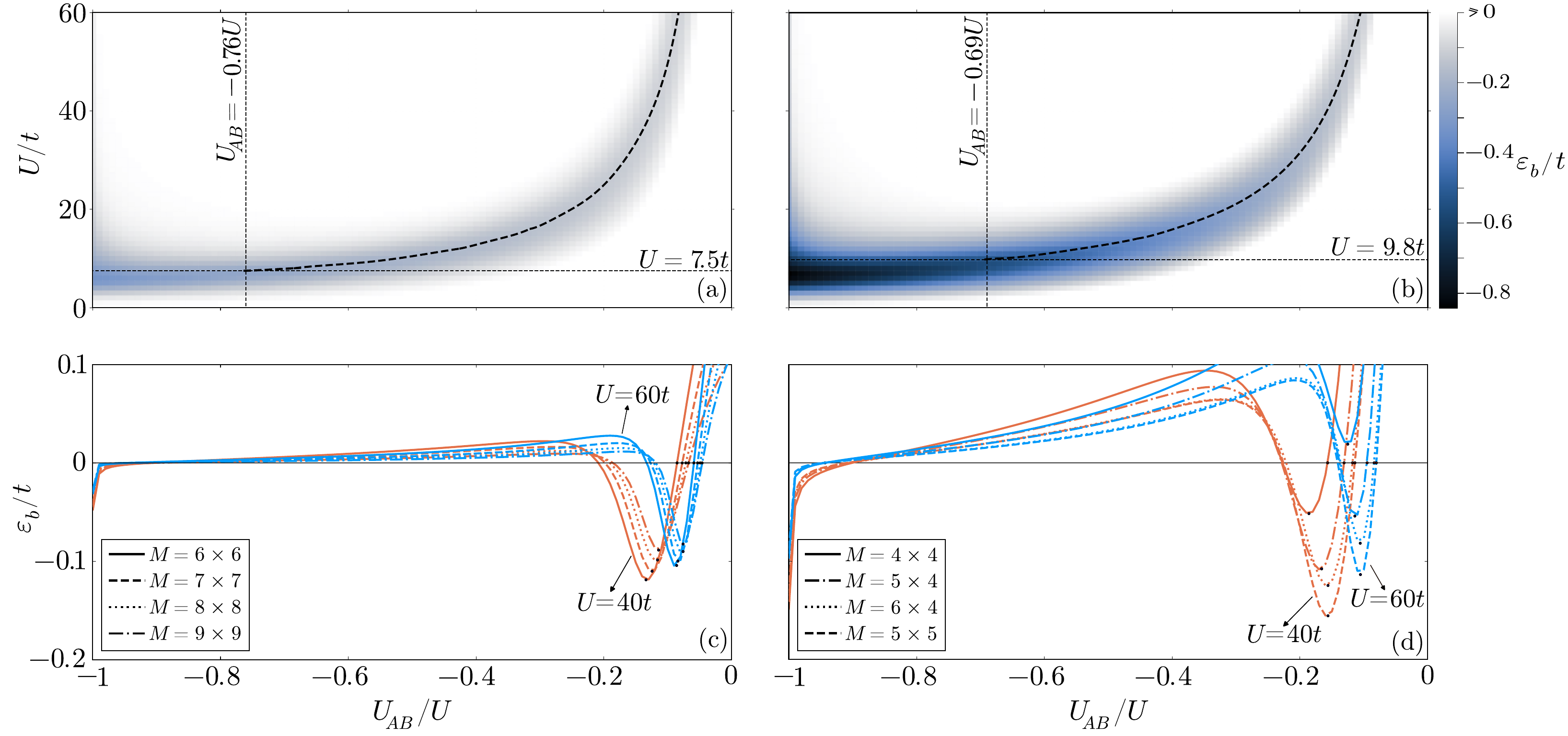}
    \caption{Ground-state binding energy $\eb$ for the AABB [(a) and (c)] and AAABBB [(b) and (d)] systems. The top panels show heatmaps of $\eb$ as a function of  $\UAB/U$ and $U/t$ for lattices with $7\times 7$ (a) and $5\times 5$ (b) sites. The dashed black lines indicate the minimum of $\eb$.
    The bottom panels show $\eb$ as a function of $\UAB/U$ for $U/t=40$ (orange lines) and $U/t=60$ (blue lines) for different lattice sizes as indicated by the legends.}
\label{sec:results;sub:eb;fig:eb}
\end{figure*}

Our two-component bosonic configuration with equal tunneling rates can be realized with one isotope in two internal states, so both components have the same mass. Such mixtures have been realized in the past, including related experiments of binary quantum droplets with $^{39}$K atoms~\cite{cabrera_quantum_2018,semeghini_self-bound_2018}. In turn, the atoms would need to be confined to a quasi-two-dimensional geometry and to an in-plane optical lattice created by ad hoc laser beams~\cite{hadzibabic_trapped_2008}. Importantly, the tight lattices considered in this work are achieved with large lattice depths, which can be controlled by the laser intensity~\cite{lewenstein_ultracold_2012}. Examples of experimental realizations of two-dimensional Bose-Hubbard models can be found in Refs.~\cite{zhang_observation_2012,cocchi_equation_2016}.

The parameters in Bose-Hubbard Hamiltonians depend non-trivially on several experimental parameters, such as recoil energies $E_R$ and scattering lengths~\cite{blakie_wannier_2004,walters_ab_2013}. In particular, the high ratios of $U/t$ considered later in this work can be achieved by controlling the lattice depths and recoil energies~\cite{blakie_wannier_2004}. In this direction, Ref.~\cite{cocchi_equation_2016} explored the regime $0\lesssim U/t\lesssim 20$ in two dimensions by employing lattice spacings of $a=532$ nm and lattice depths of $~6E_R$. Magnetic fields around 200 G were used to control the interactions, which could be tuned further.

\section{Results}
\label{sec:results}

In this section, we examine the main properties that characterize the formation of bound states in the ground state of the system. An examination of excited states is provided in App.~\ref{app:spectrum} to explore the stability of the ground state and the structure of the excited states.  Additionally, imbalanced configurations are examined in App.~\ref{app:imbalanced}, complementing the results of the main text on balanced configurations.

\subsection{Binding energies}
\label{sec:results;sub:eb}

We start by examining the binding energies $\eb$ to elucidate the formation of bound states. The binding energies are calculated by the minimal difference between the ground-state energy of the full system and that of its components~\cite{guijarro_few-body_2020}. For the four-body system, this corresponds to
\begin{equation}
    \eb=E_{AABB}-2E_{AB},
\label{sec:results;sub:eb;eq:ebAABB}
\end{equation}
whereas for the six-body one to
\begin{equation}
\eb=E_{AAABBB}-\min(E_{AABB}+E_{AB},3E_{AB}),
\label{sec:results;sub:eb;eq:ebAAABBB}
\end{equation}
where $E$ are the ground-state energies of the system indicated by the subscript. Eq.~(\ref{sec:results;sub:eb;eq:ebAABB}) essentially compares whether the formation of one bound tetramer is favorable to the formation of two independent dimers.
Similarly, Eq.~(\ref{sec:results;sub:eb;eq:ebAAABBB}) compares whether the formation of a bound hexamer is favorable to that of either three dimers or one tetramer and dimer. Therefore, a negative binding energy ($\eb<0$) indicates the formation of a bound tetramer or hexamer in the four- or six-body system, respectively. 

Here we note that, for the AABB system, one does not need to compare with terms such as $E_{AAB}+E_B$, as having a free particle is never favorable (also see App.~\ref{app:imbalanced} for details on the AAB system). Similarly, for the AAABBB system, we do not need to compare with $E_{AA}$ or $E_{BB}$, as bound states of only atoms of the same species are not possible due to the intra-species repulsion.

We show a representative set of binding energies in Fig.~\ref{sec:results;sub:eb;fig:eb}. The left and right panels show results for the four- and six-body systems, respectively.  The top panels show heatmaps for a wide range of interspecies and intraspecies interactions for fixed lattice sizes. In turn, the bottom panels show curves for selected values of $U/t$ and different lattice sizes to examine the dependence of the results on the number of sites. 

The top panels show that $\eb$ becomes negative for a wide range of interactions, indicating that bound clusters of all available bosons are indeed formed for $|\UAB|/U<1$ and $U/t\gtrsim 5$. In such cases, the bound states are formed within a range $\UAB^{(1)}<U<\UAB^{(0)}$, where $\UAB^{(1)}$ and $\UAB^{(0)}$ are the critical interaction strengths where $\eb=0$. We have found qualitatively similar results for different lattice sizes. This is illustrated by the bottom panels, where $\eb$ shows a similar behavior for different choices of $M$. However, a clear convergence pattern is difficult to observe, as ED calculations typically display a very slow convergence with the number of sites~\cite{raventos_cold_2017}. Moreover, no hexamer is found for $M=4\times 4$ and $U=60t$ [see panel (d)], which we attribute to the small lattice size.

One striking feature of our results is that the binding energy shows a local minimum $\eb^*$ at an interaction strength $\UAB^*$ for certain values of $U/t$. In the top panels, this minimum is indicated by the dashed black lines. We have found that the minimum appears at intraspecies interaction strengths of $U/t\gtrsim 7.5$ (AABB) and $U/t\gtrsim 9.8$ (AAABBB), corresponding to small tunneling rates (large interactions). However, we note that this minimum could appear for larger tunneling rates in larger lattices.

The appearance of $\eb^*$ is better illustrated by the bottom panels, where it appears for all lattice sizes. This minimum is a non-trivial quantum effect, as one would expect that an increasing interspecies attraction $\UAB$ would simply increase the magnitude of the binding energy. Instead, $\eb$ shows a non-monotonic dependence on $\UAB$. The binding energy only shows a monotonic decrease with $\UAB$ in the collapsing region $\UAB/U<-1$ for these large values of $U/t$.  

Here we mention that, in contrast, in one-dimensional chains the binding energy does show a monotonic decrease with $\UAB$ for all interactions~\cite{morera_universal_2021} (see App.~\ref{app:1D} for supporting results). We have checked that the non-monotonic dependence of $\eb$ on $\UAB$ is present when the coordination number is greater than two. Therefore, we can conclude that the behavior of clusters in one-dimensional chains is very different from that in higher-dimensional lattices.

The non-monotonic dependence of $\eb$ on $\UAB/U$ is nonetheless similar to the behavior found in the two-dimensional continuum (see Ref.~\cite{guijarro_few-body_2020} for a comprehensive examination). In such a case, $\eb$ shows a minimum in $a_{AB}/a$, with $a_{AB}$ and $a$ the interspecies and intraspecies scattering lengths, respectively. There, the scattering lengths play a similar role to the interaction strengths $U$ and $\UAB$, even though a direct comparison is not possible due to the additional dependence on the lattice potential~\cite{blakie_wannier_2004}. However, in the continuum, the binding energies are universal functions of $a_{AB}/a$, and the bound states are formed for any value of $a$. In contrast, in the small lattice systems examined here, $\eb$ does not show any universal behavior, having a different dependence on $\UAB/U$ for different values of $U/t$. In turn, as already stressed, the bound states are formed beyond a strong repulsion $U/t\gtrsim 5$. Moreover, in our calculations, both the magnitude of $\eb^*$ and the minimum's position $\UAB^*$ change with $U/t$.

Here we note that, for large values of $U/t$, the tunneling is heavily suppressed. Therefore, in such a regime, we can expect the lattice system to behave very differently from the continuum due to the enhanced effects of the lattice potential. However, a comprehensive comparison with the continuum should be performed in the future by studying larger lattices with other approaches.

\begin{figure}
    \centering \includegraphics[width=\columnwidth]{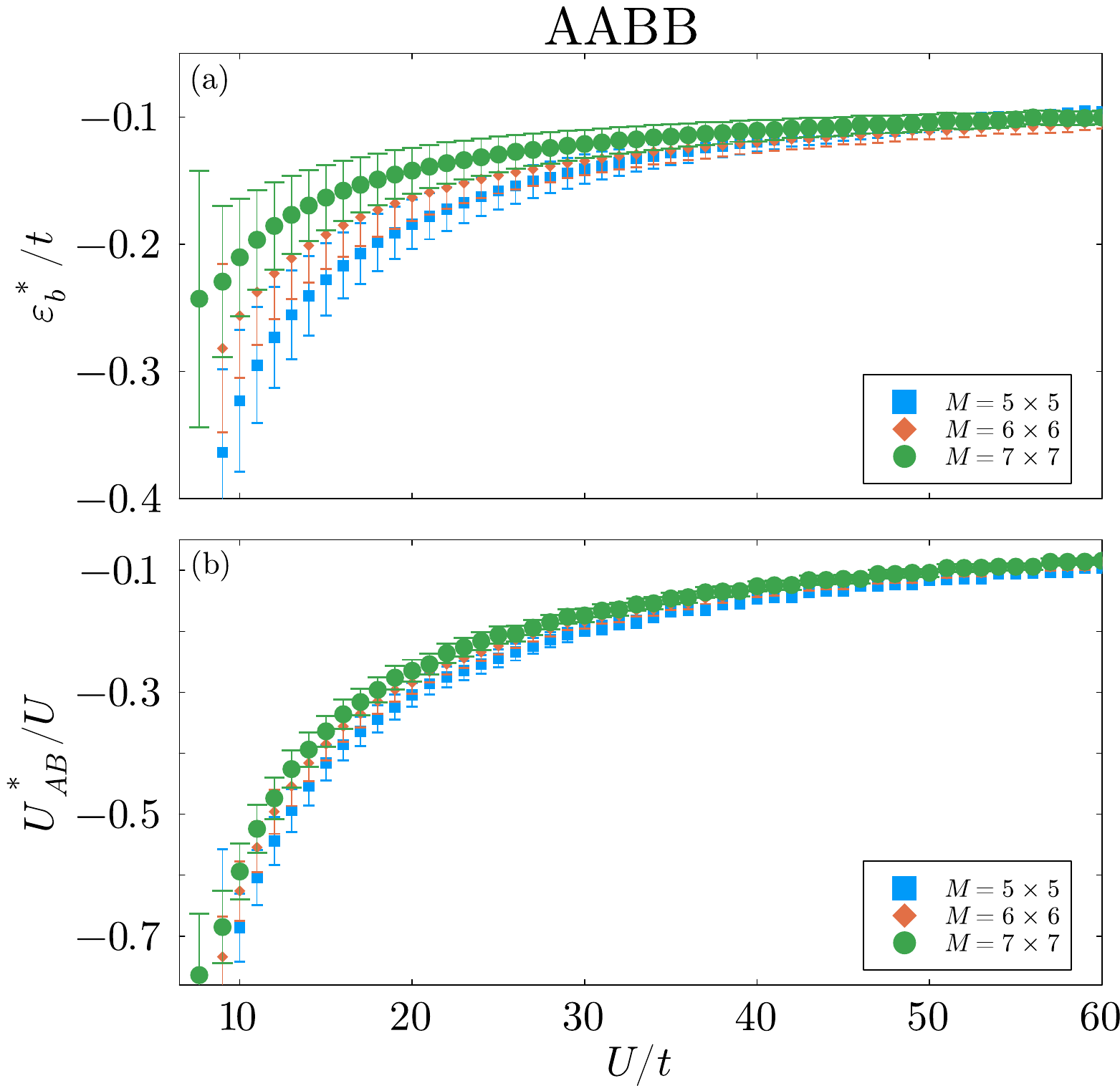}
    \caption{Magnitude of the local minimum of the binding energy $\eb^*$ (a) and the interaction strength at which it appears $\UAB^*/U$ (b) for the AABB system as a function of $U/t$. The minima are located with a golden-section search algorithm, and thus the error bars indicate the associated error.
    Results for different lattice sizes are reported, as indicated by the legends. }
    \label{sec:results;sub:eb;fig:min}
\end{figure}

Finally, to illustrate how the value of $\eb^*$ and its position $\UAB^*$ change, in Fig.~\ref{sec:results;sub:eb;fig:min} we show these quantities as a function of $U$ for the four-body system. From panel (a), we observe that $\eb^*$ increases with $U$ for all lattice sizes. The values for different lattice sizes show noticeable differences for $U/t \lesssim 40$. However, for very large $U$, the binding energy converges to $\eb^*\approx -0.1 t$ in all cases.

Panel (b) shows how $\UAB^*$ also increases with $U$.
Importantly, for all the lattice sizes examined, the position of the minimum also converges to $\UAB^*/U\approx -0.1$ for very large $U/t$. At these large interactions, the tunneling rate plays a negligible role. Therefore, the only remaining scale in the system is $\UAB/U$.
Overall, our results seem consistent across the lattice sizes examined in this work.

\subsection{Average distances between atoms}
\label{sec:results;sub:distances}

\begin{figure*}[t]
    \centering
    {\includegraphics[width=\textwidth]{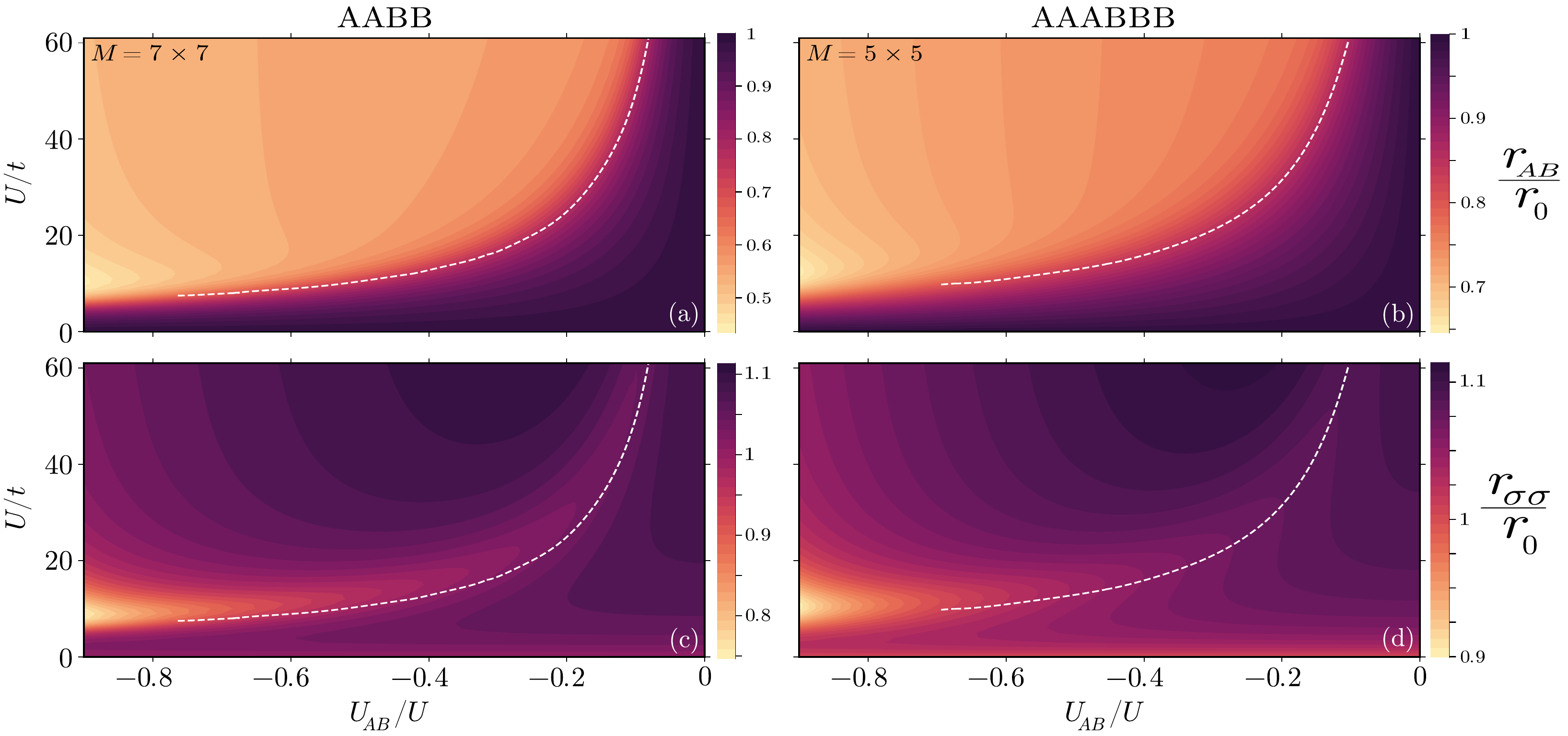}}
    \caption{Average distances for the AABB [(a) and (c)] and AAABBB [(b) and (d)] systems as a function of  $\UAB/U$ and $U/t$. The top panels show distances between the two different species $r_{AB}$, while the bottom panels show distances between the same species $r_{\sigma\sigma}$. The four- and six-body systems consider lattices with $7\times 7$ and $5\times 5$ sites, respectively.
    The average distance between two non-interacting bosons, $r_0$, is $\approx 2.859d$ for the $7\times 7$ lattice, while it is $\approx 2.009d$ for the $5\times 5$ lattice. The dashed white lines indicate the position of the minimum of $\eb$.}
\label{sec:results;sub:distances;fig:distances}
\end{figure*}

To further understand the behavior of the bound states, one can examine the distribution of the particles across the lattice. Because our system is periodic, we cannot employ the density profiles, as the profiles are simply a constant number. Therefore, we instead examine the average distance between the atoms. Within our ED calculations, this is calculated by computing the distance between bosons in each Fock state~\cite{isaule_bound_2024}. In a square lattice, this takes the form:
\begin{equation}
    r_{\sigma\sigma'}= \frac{d}{\mathcal{N}} \sum_{\nu=1}^\mathcal{D} \sum_{i,j}|c_\nu|^2 n_{i,\sigma}^{(\nu)}n_{j,\sigma'}^{(\nu)}r(i,j),
\end{equation}
where $d$ is the lattice spacing, $\mathcal{N}$ is the number of distances to count, and $r(i,j)$ gives the distance between two sites $i$ and $j$. The second sum runs over all the sites $i$ and $j$. For the lattices considered in this work, the distances are given by~\cite{isaule_bound_2024}
\begin{align}
    r(i,j)^2 =& \Big( \min(|i_x - j_x|, M_x - |i_x - j_x|) \Big)^2\nonumber \\
    &+\Big( \min(|i_y - j_y|, M_y - |i_y - j_y|) \Big)^2,
\end{align}
where $i=(i_x,i_y)$ and $j=(j_x,j_y)$ give the $x$- and $y$-coordinates of each site. In this formula, the $\min$-functions are included to take the periodicity of the lattices into account. Note that $\mathcal{N}=N_A N_B$ for distances between bosons of different species, while $\mathcal{N}=\binom{N_\sigma}{2}$ for distances between the same species, where $\binom{.}{.}$ is the binomial coefficient.

In Fig.~\ref{sec:results;sub:distances;fig:distances} we show the average distance between species $A$ and $B$ (top panels) and between the same species (bottom panels). As in Fig.~\ref{sec:results;sub:eb;fig:eb}, the left panels show results for the four-body system, while the right panels show results for the six-body system. All the distances are rescaled in terms of the distance between two non-interacting bosons $r_0$ for the respective lattice sizes. We show the results as contour maps, as they better highlight the regions where the distances change. Again, similar results are obtained for the four- and six-body systems.

Firstly, we observe that the distance $r_{AB}$ between different species (top panels) is always lower than $r_0$. This is expected, as the distance between $A$ and $B$ must decrease when bound states are formed. Therefore, $r_{AB}$ decreases with decreasing $\UAB$. More importantly, $r_{AB}$ seems strongly correlated with the behavior of the binding energy $\eb$. The distance $r_{AB}$ is approximately constant along the local minimum of $\eb$. The latter is indicated by the white dashed line where $r_{AB}$ approximately follows the same contour.
Additionally, we find that $r_{AB}$ changes abruptly around $\UAB^*$ (white dashed line). For $\UAB>\UAB^*$ we find that $r_{AB}\approx r_0$, while for $\UAB<\UAB^*$ the distance rapidly decreases to values around  $r_{AB}\approx 0.5r_0$ for the tetramer and  $r_{AB}\approx 0.7r_0$ for the hexamer.

Secondly, the distance $r_{\sigma\sigma}$ between bosons of the same species (bottom panels) does not show a monotonic decrease with $\UAB$, in contrast to $r_{AB}$. In particular, we find that $r_{\sigma\sigma}$ also behaves very differently around $\UAB^*$. As also indicated by the white dashed lines, $r_{\sigma\sigma}$ reaches a local minimum approximately at $\UAB^*$. Indeed, the peaks of the different contours noticeably follow the white dashed lines. The difference between the behavior of $r_{AB}$ and $r_{\sigma\sigma}$ is explained by the competition between the attractive interspecies and the repulsive intraspecies interactions. Intuitively, two bosons of different species prefer to localize at the same site due to the attraction $\UAB<0$, explaining the smaller values of $r_{AB}$. On the other hand, two bosons of the same species prefer to localize at different sites due to the repulsion $U>0$, and thus $r_{\sigma\sigma}$ is larger. However, because bound clusters are still formed, the distance $r_{\sigma\sigma}$ still decreases when the bound states are formed. Overall, we have checked that qualitatively similar results are found with other lattice sizes.

\begin{figure}
    \centering \includegraphics[width=\columnwidth]{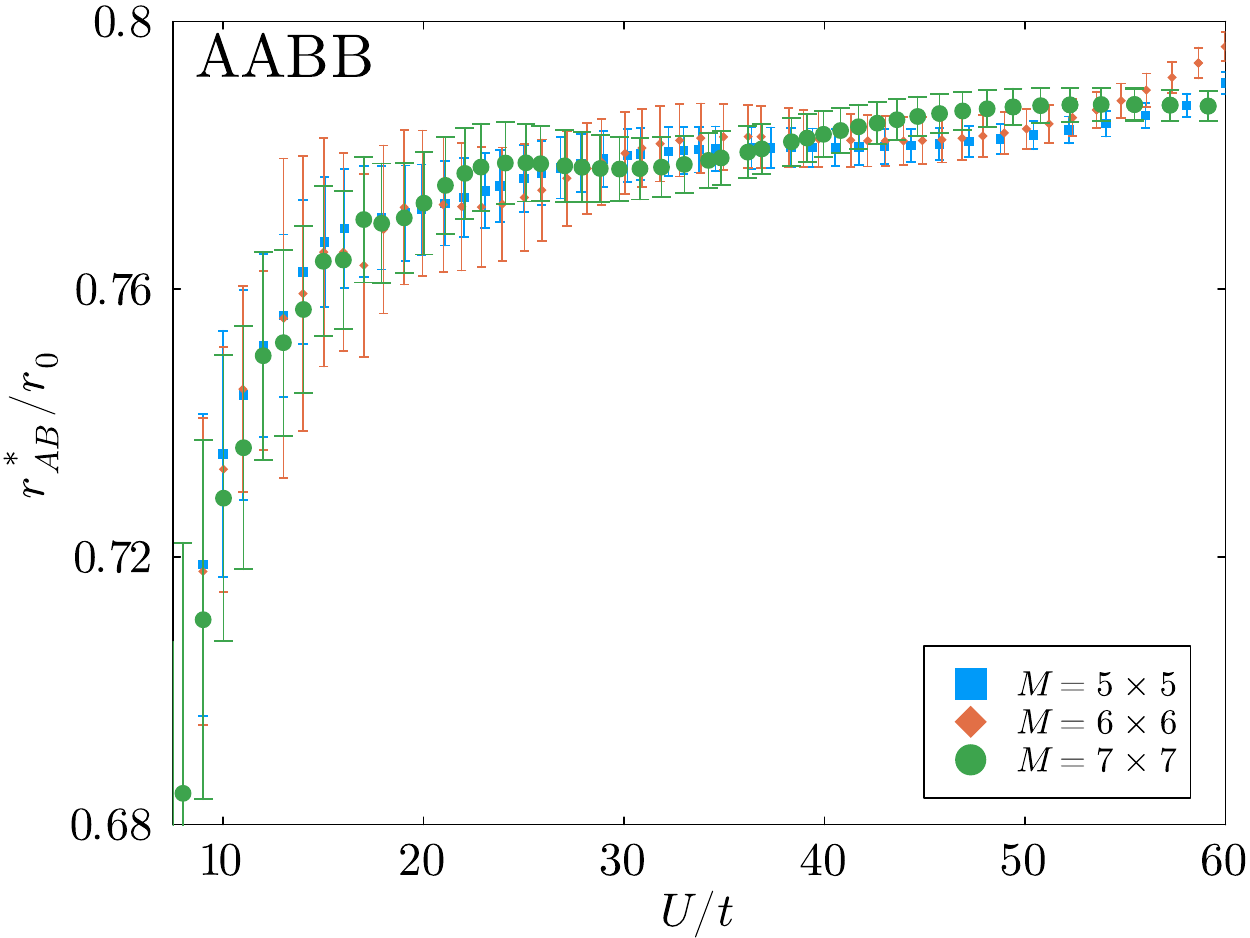}
    \caption{Interspecies distance $r_{AB}$ at $\UAB^*$ for the AABB system as a function of $U/t$. The minima are located with a golden-section search algorithm, and thus the error bars indicate the associated error. Results for different lattice sizes are reported, as indicated by the legend.}
    \label{sec:results;sub:distances;fig:rc}
\end{figure}

Finally, to better illustrate how the interspecies distance behaves at the minimum of the binding energy, in Fig.~\ref{sec:results;sub:distances;fig:rc} we show the value of $r_{AB}$ at $\UAB^*$ as a function of $U/t$ for the four-body system. We refer to this distance as $r_{AB}^*$. We observe that $r_{AB}^*$ shows a noticeable dependence on $U$ for $U/t\lesssim 20$ (left side of the figure), where it increases from $\approx 0.65r_0$ to $\approx 0.78 r_0$. However, the distance becomes approximately constant for stronger intraspecies interactions $U/t\gtrsim 20$, confirming that $r_{AB}^*$ follows the same contour in Fig.~\ref{sec:results;sub:distances;fig:distances}(a). This saturation of $r_{AB}^*$ to a constant value for large $U/t$ is also consistent with the saturation reported in Fig.~\ref{sec:results;sub:eb;fig:min}, where the tunneling rate does not play a role.

Importantly, Fig.~\ref{sec:results;sub:distances;fig:rc} shows that lattices with different sizes result in similar values of $r_{AB}^*$ across different interaction strengths. We stress that this invariance within different lattice sizes can only be observed by rescaling the distances by $r_0$. In addition, we note that the apparent oscillations of $r_{AB}^*$ are an artifact of the algorithm used to locate the minima.

\subsection{Entanglement}
\label{sec:results;sub:S}

\begin{figure*}[t]
    \centering
    \includegraphics[width=\textwidth]{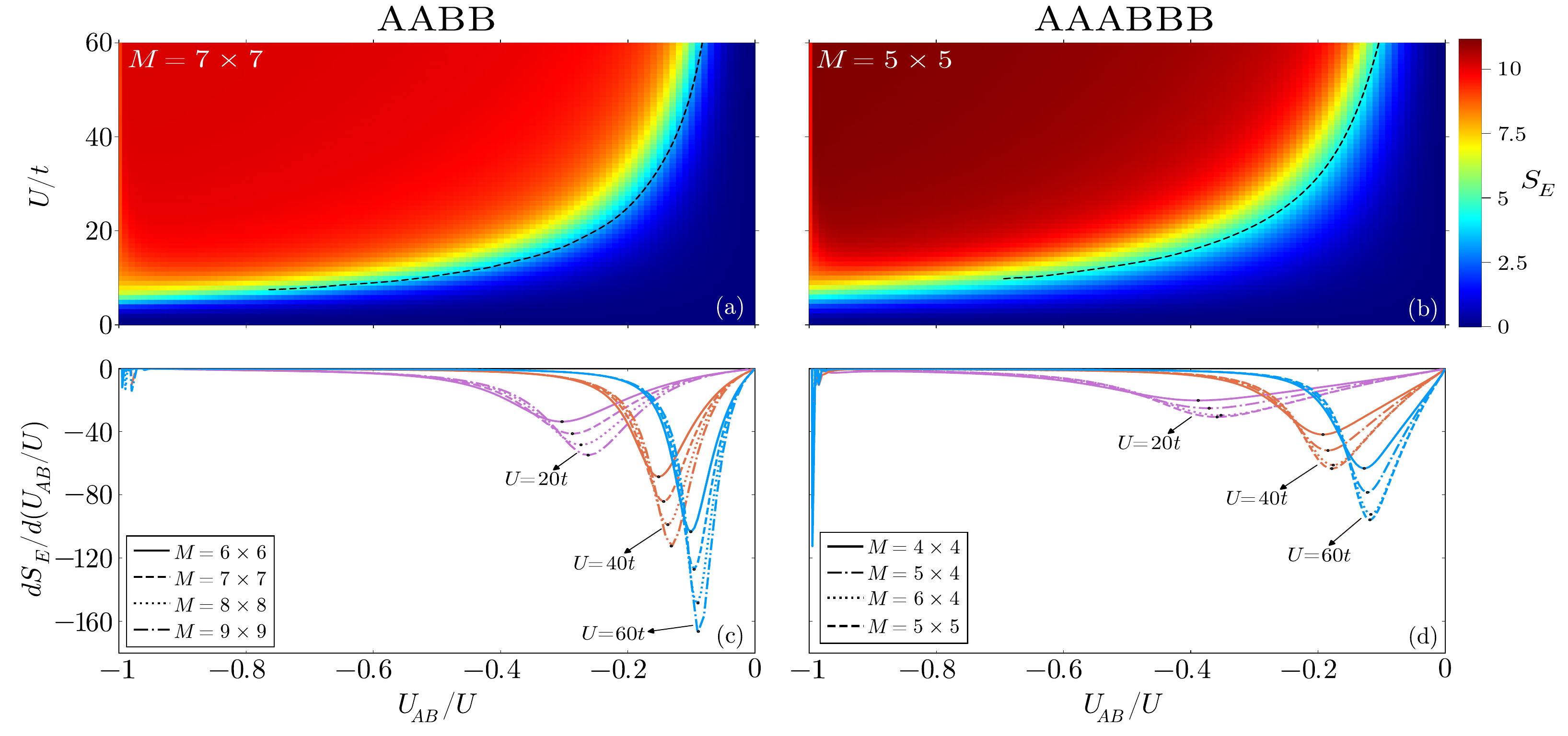}
    \caption{Von Neumann entropy $S_E$ (top panels) and its derivative with respect to $\UAB/U$ (bottom panels)  for the AABB [(a) and (c)] and AAABBB [(b) and (d)].  The top panels show heatmaps of $S_E$ as a function of  $\UAB/U$ and $U/t$ for lattices with $7\times 7$ (a) and $5\times 5$ (b) sites. The dashed black lines indicate the minimum of $\eb$. The bottom panels show $dS_E / d(U_{AB}/U)$ as a function of $\UAB/U$ for $U/t=20$ (purple lines), $U/t=40$ (orange lines), and $U/t=60$ (blue lines) for different lattice sizes as indicated by the legends.}
    \label{sec:results;sub:S:fig:S}
\end{figure*}

The formation of bound states means that species $A$ and $B$ become entangled. To measure the entanglement between the two species, we compute the von Neumann entropy $S_E$. Among many applications, this entropy has been used in the characterization of quantum phase transitions~\cite{amico_entanglement_2008}, as well as in the formation of two-particle bound states~\cite{brand_localization_2008}. We refer to Refs.~\cite{mujal_quantum_2016,richaud_pathway_2019} for related applications in Bose-Bose mixtures. The von Neumann entropy reads
\begin{equation}
    S_E = -\tr[\rho_\sigma \log_2\,\rho_\sigma] = -\sum_{k=1}^{\mathcal{D}_{\sigma}} \lambda^{(\sigma)}_k \log_2\,\lambda^{(\sigma)}_k,
\label{sec:results;sub:S;eq:S}
\end{equation}
where $\rho_\sigma$ is the reduced density matrix of the $\sigma=A,B$ sub-system and $\lambda_k$ are the eigenvalues obtained from its singular value decomposition (SVD)
\begin{equation}
    \rho_\sigma=\sum_{k=1}^{\mathcal{D}_{\sigma}}\lambda_k^{(\sigma)}|\lambda_k^{(\sigma)}\rangle\langle \lambda_k^{(\sigma)}|.
\end{equation}
The entropy takes values $0\leq S_E \leq \log_2(\min(\mathcal{D}_A,\mathcal{D}_B))$. It vanishes ($S_E=0$) when there is no entanglement between the species, while it reaches its maximum when the species are maximally entangled. This maximum value is obtained by imposing the condition of maximum entanglement, $\lambda_k^{(\sigma)}=1/\min(\mathcal{D}_A,\mathcal{D}_B)$ for \emph{all} $k$, into Eq.~(\ref{sec:results;sub:S;eq:S}) (see, for example, Ref.~\cite{mujal_quantum_2016}).

The top panels of Fig.~\ref{sec:results;sub:S:fig:S} show the von Neumann entropy $S_E$ as a function of the interspecies and intraspecies interactions. As with previous examined quantities, both the four- and six-body systems show similar results. It is evident that $S_E$ changes abruptly around $\UAB^*$, the minimum of the binding energy, further confirming that such a minimum is a special point in the system. For $\UAB>\UAB^*$ (lower right regions of the panels), the entropy is approximately zero, showing that the two species are essentially not entangled. In contrast, for $\UAB<\UAB^*$ (upper left regions of the panels), the entropy quickly reaches a constant maximum entropy of $S_E\approx 10.17$ and $S_E\approx 11.14$ for the four- and six-body systems, respectively. The latter means that the system becomes maximally entangled for interactions $\UAB<\UAB^*$.

To illustrate how $S_E$ changes around $\UAB^*$, we show the derivative of the entropy with respect to the interspecies interactions in the bottom panels of Fig.~\ref{sec:results;sub:S:fig:S}. For readability, the figure shows derivatives for selected choices of $U$. In all cases, the magnitude of the derivative greatly increases around $\UAB^*$, while it vanishes elsewhere. This confirms that $S_E$ only changes around the minimum of the binding energy, where the species quickly become entangled. Therefore, the von Neumann entropy can be used to measure the formation of the bound states.

\begin{figure}[t!]
    \centering \includegraphics[width=\columnwidth]{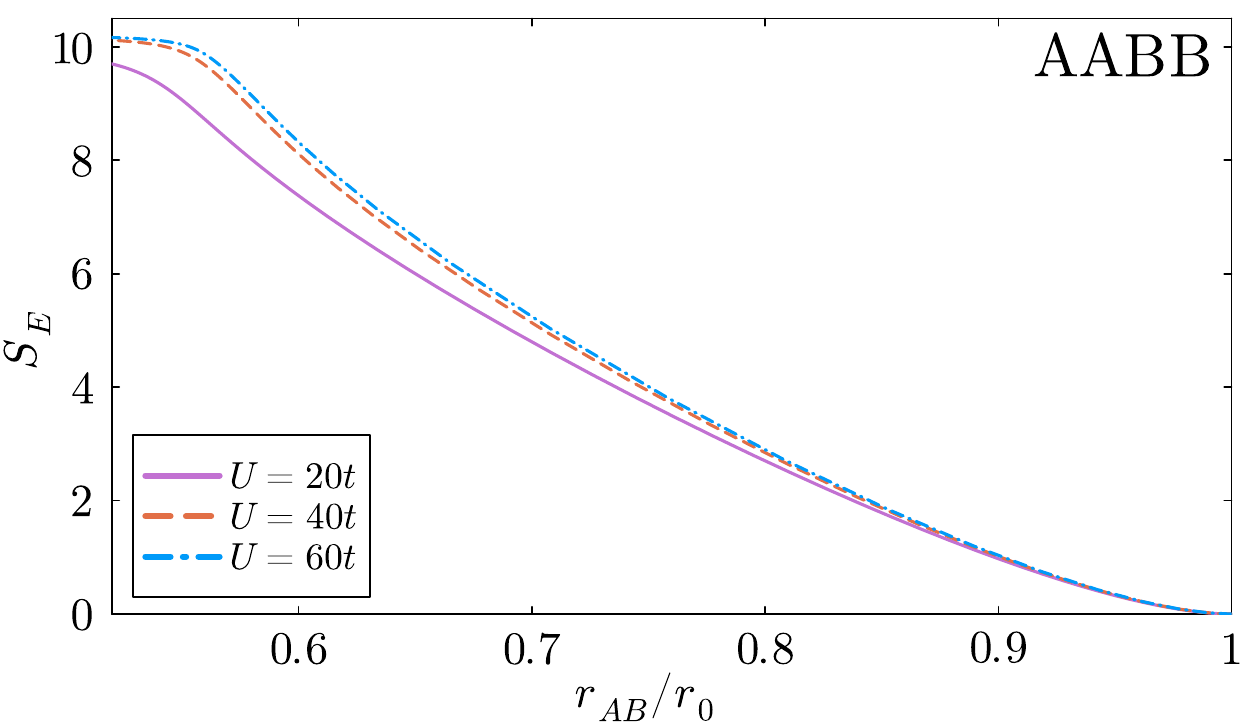}
    \caption{Von Neumann entropy $S_E$ as a function of the interspecies distance $r_{AB}$ for the AABB system.  The curves show results for $U/t=20$ (purple lines), $U/t=40$ (orange lines), and $U/t=60$ (blue lines). The figure considers a lattice with $7\times7$ sites.}
    \label{sec:results;sub:S:fig:SoR}
\end{figure}

We note that, unlike other quantities, the magnitude of $S_E$ and its derivative change strongly when increasing the number of sites. Moreover, the values of the entropy show a noticeable difference between the four- and six-body systems. This can be expected, as the magnitude of $S_E$ depends on the size of the lattice of particles. However, the relevant physical information is given by the change of the entropy, which is consistent in all the cases examined.

Finally, in Fig.~\ref{sec:results;sub:S:fig:SoR} we show the von Neumann entropy as a function of the inter-species distance. We observe that the entropy vanishes in the limit of $r_{AB}=r_0$ (bottom right corner), confirming that there is no entanglement when the distance is that of free particles. Expectedly, the entropy shows a monotonic increase with decreasing $r_{AB}$. We find that the distance between species becomes minimal when the von Neumann entropy reaches its maximum value (top left corner), corresponding to the point of largest entanglement.

\section{Conclusions}
\label{sec:concl}

In this work, we have examined the formation of bound states in systems with a binary mixture of a few bosons in small square optical lattices. We have shown that bound clusters composed of all available bosons are formed for intermediate interspecies interactions. Importantly, we have found that, for small tunneling rates, the binding energy has a non-monotonic dependence on the interspecies interaction strength, developing a local minimum. This non-monotonic dependence does not appear in one-dimensional lattices as studied in the past, showing that the binding mechanics of binary bosonic mixtures in optical lattices depend on the lattice geometry. In addition, while the non-monotonic behavior of the binding energy also appears in the two-dimensional continuum, in that case, the properties of the system are universal, and the bound states appear for any intraspecies repulsion. In contrast, in the small lattice systems examined in this work, the properties depend on the physical parameters, and the bound states only appear for large intra-species interactions.

The minimum of the binding energy is also a special point in the system, as different physical properties change around the minimum. The average distance between different species is approximately constant for different tunneling rates along the binding energy's minimum. Interestingly, entanglement properties are very sensitive to the minimum, as the von Neumann entropy changes abruptly around that point. 

The studied bound states can be considered as a precursor of many-body droplet states in binary bosonic mixtures. Therefore, the reported results suggest that quantum droplets in two- and three-dimensional lattices could have a distinct dependence on the interatomic interaction from the one-dimensional droplets studied in the past~\cite{morera_quantum_2020,morera_universal_2021}. A particularly relevant extension of the present work would be to find the connection between the found minimum of the binding energy and the equilibrium condition of large droplets. Such studies could be performed in the future with other numerical and theoretical techniques, such as quantum Monte-Carlo~\cite{gubernatis_quantum_2016,motoyama_dsqss_2021} or the Quantum Gutzwiller approach~\cite{caleffi_quantum_2020,colussi_quantum_2022}. Such approaches could also be employed to test our results in larger lattices, which is a pressing issue. 

\begin{acknowledgments}
We thank Bruno Juliá-Díaz for the helpful discussions. M.V.A. was supported by Vicerrectoría de Investigación UC. F.I. acknowledges funding from ANID through FONDECYT Postdoctorado No. 3230023. 

\end{acknowledgments}

\section*{Data Availability}

The data that support the findings of this article are openly available~\footnote{Data for: Data for: Few-body bound states of bosonic mixtures in two-dimensional optical lattices (2025), Zenodo, \href{http://doi.org/10.5281/zenodo.15775764}{10.5281/zenodo.15775764}.}.

\appendix

\section{Low-energy spectrum}
\label{app:spectrum}

\begin{figure}[t!]
    \centering
    \includegraphics[width=\columnwidth]{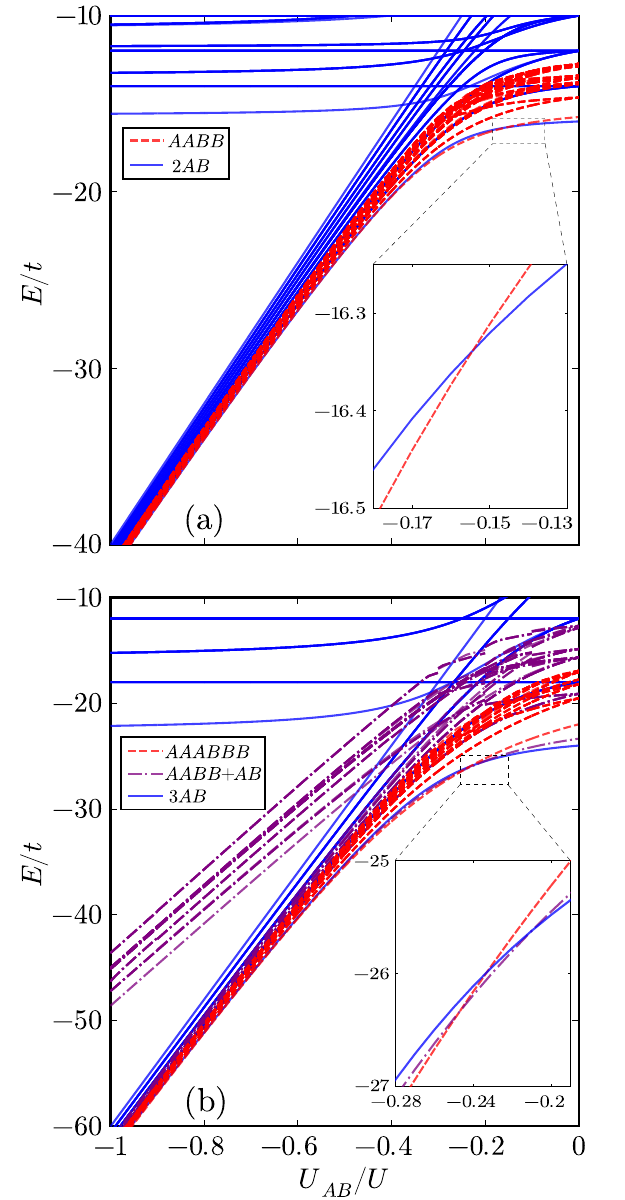}
    \caption{Low energy spectrum for $U = 40t$ for the four- (a) and six-body (b) system as a function of $\UAB/U$. $(a)$. The specific configurations are indicated by the legends. Panel (a) considers lattices with $6\times 6$ sites and shows the 100 lowest energy levels, while panel (b) considers lattices with $4\times 4$ sites and shows the 50 lowest energy levels. The insets show the crossing of the ground states.}
\label{app:spectrum;fig:spectrum}
\end{figure}

\begin{figure*}[t!]
    \centering
    {\includegraphics[width=\textwidth]{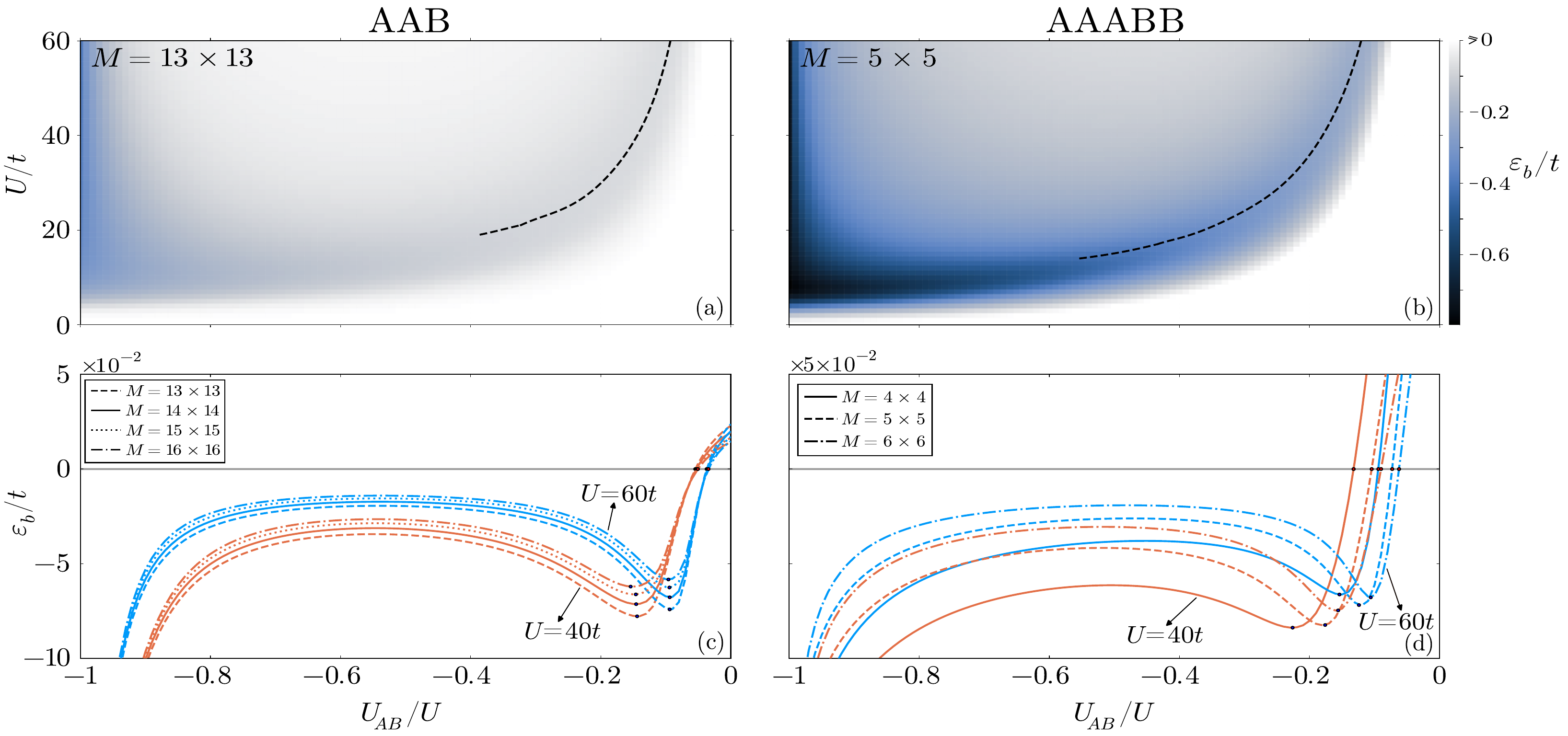}}
    \caption{Ground-state binding energy $\eb$ for the AAB [(a) and (c)] and AAABB [(b) and (d)] systems. The top panels show heatmaps of $\eb$ as a function of  $\UAB/U$ and $U/t$ for lattices with $13\times 13$ (a) and $5\times 5$ (b) sites. The dashed black lines indicate the minimum of $\eb$.
    The bottom panels show $\eb$ as a function of $\UAB/U$ for $U/t=40$ (orange lines) and $U/t=60$ (blue lines) for different lattice sizes as indicated by the legends.}
    \label{app:imbalanced;fig:eb}
\end{figure*}

While this work focuses on ground-state properties, examining the low-energy spectrum can give us insight into the formation of energy bands of different configurations (see, for example, Ref.~\cite{pohlmann_trion_2013}). In addition, it enables us to estimate the stability of the bound states. Thus, we show the low-energy spectrum for the examined balanced systems in Fig.~\ref{app:spectrum;fig:spectrum}. The panels show the spectrum of the full AABB [panel (a)] and AAABBB [panel (b)] systems, as well as their sub-systems.

Panel (a) displays well-defined energy bands for the tetramer (AABB) and the two-dimer (2AB) configurations. The bands decrease with $\UAB$. The inset shows the crossing of the ground state. When the ground state of the AABB system is lower than the 2AB system, then $\eb$ becomes negative and a ground-state tetramer is formed [see Eq.~(\ref{sec:results;sub:eb;eq:ebAABB})]. However, it is easy to see that the AABB and 2AB bands appear at similar energies, showing several crossings. Therefore, we can conclude that the tetramer will break easily into two dimers, making its potential experimental observation a challenge. Nevertheless, larger energy gaps could appear in larger lattices, and thus an examination of large lattices is required to provide a definitive answer about the stability of the bound clusters.

Similarly, panel (b) displays well-defined energy gaps for the different configurations. However, the lower energy band of the AABB+AB system appears at noticeably larger energies than that of the AAABBB and 3AB systems. Therefore, the hexamer will break easily into three dimers, but not into a tetramer and a dimer.

\section{Imbalanced mixtures}
\label{app:imbalanced}

In the following, we examine the onset of bound states in imbalanced mixtures to demonstrate that the reported results also hold when there is an imbalance in the species populations. Here we examine the cases when $N_A=2$ and $N_B=1$ (AAB), 
and when $N_A=3$ and $N_B=2$ (AAABB). Nevertheless, we have checked that other imbalanced mixtures, such as the imbalanced tetramer AAAB, show similar results. 

The binding energy $\eb$ of a trimer in the AAB system reads
\begin{equation}
    \eb=E_{AAB}-E_{AB}-E_{A},
\label{sec:app;sub:eb;eq:ebAAB}
\end{equation}
whereas that of a pentamer in the AAABB system reads
\begin{equation}
    \eb=E_{AAABB}-E_{AAB}-E_{AB}.
\label{sec:app;sub:eb;eq:ebAAABB}
\end{equation}
Therefore, Eq.~(\ref{sec:app;sub:eb;eq:ebAAB}) compares whether the formation of a bound trimer is favourable to the formation of an AB dimer with a free unbound boson. Similarly, Eq.~(\ref{sec:app;sub:eb;eq:ebAAABB}) compares whether the formation of a bound pentamer is favourable to the formation of one trimer and dimer.

We show binding energies for the two examined imbalanced configurations in Fig.~\ref{app:imbalanced;fig:eb}. We use the same panel figure convention as in Sec.~\ref{sec:results;sub:eb}. Overall, the binding energy behaves similarly to the balanced cases. The binding energy becomes negative, signalling the formation of bound states. The binding energy also shows a non-monotonic dependence on $\UAB$ with a local minimum at an interaction strength $\UAB^*$ for small tunneling rates. Following previous conventions, the minimum is indicated by the black dashed lines in the top panels, which can also be appreciated in greater detail in the bottom panels.

Note that Fig.~\ref{app:imbalanced;fig:eb}(b) shows that, in the AAABB system, the binding energy changes somewhat strongly with the lattice size. Nevertheless, we can conclude that the formation of bound states in imbalanced mixtures is robust, as we have also checked that bound clusters are formed in imbalanced four- and six-body systems.  Further examinations of imbalanced mixtures could be explored in greater detail in the future. In particular, the polaron and bipolaron limit could be investigated and contrasted against the one-dimensional case~\cite{isaule_bound_2024}.

\section{One-dimensional lattices}
\label{app:1D}

\begin{figure}[t!]
    \centering
    {\includegraphics[width=\columnwidth]{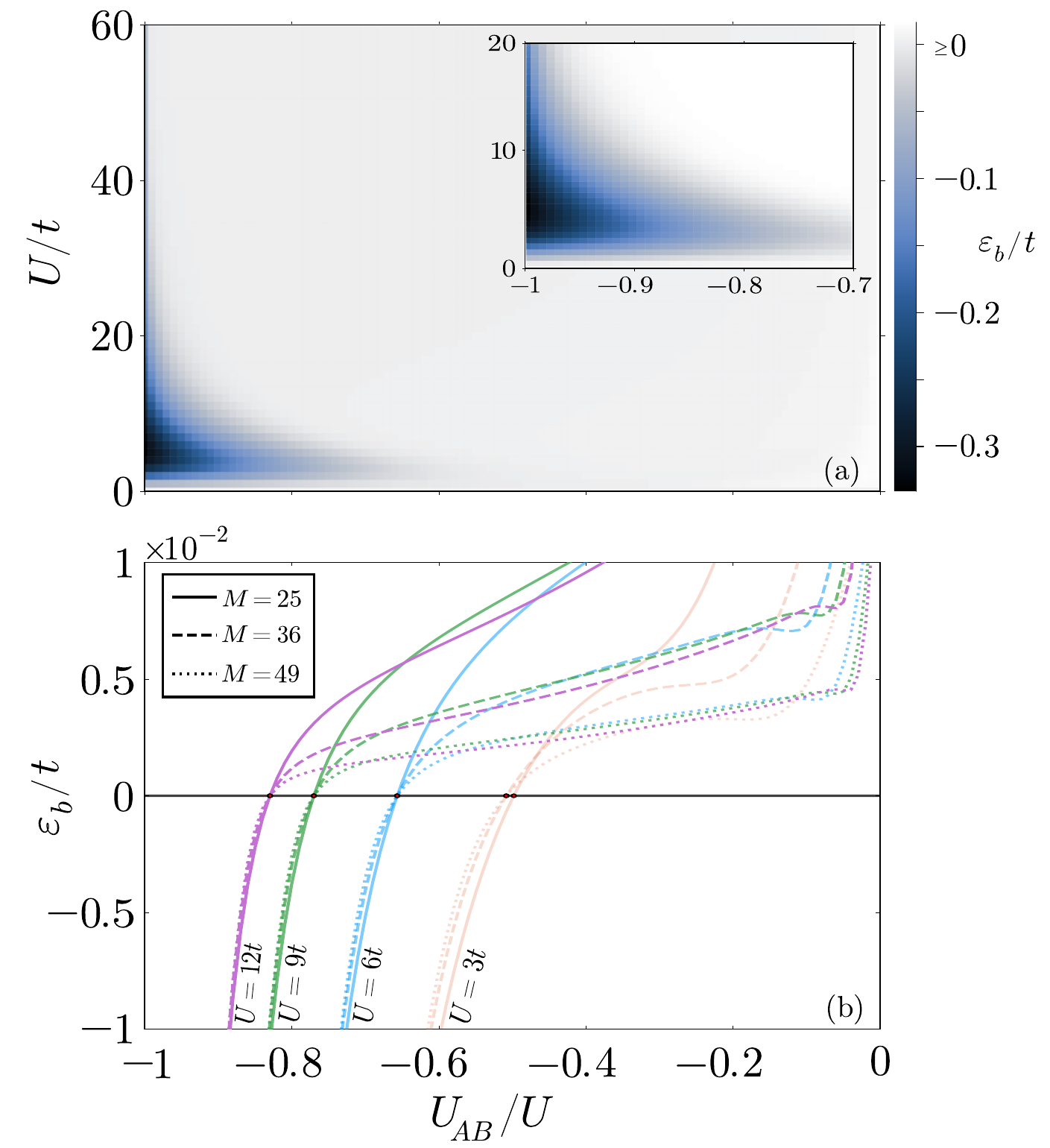}}
    \caption{Ground-state binding energy $\eb$ for the AABB system. Panel (a) shows a heatmap of $\eb$ as a function of  $\UAB/U$ and $U/t$ for a lattice with 49 sites. Panel (b) shows $\eb$ as a function of $\UAB/U$ for $U = 3t$ (orange curves), $U = 6t$ (blue curves), $U = 9t$ (green curves), and $U = 12t$ (purple curves) for different lattice sizes as indicated by the legend. }
    \label{app:1d;fig:eb}
\end{figure}

As mentioned in the main text, while bound states are also formed in one-dimensional chains~\cite{morera_universal_2021}, they show a monotonic decrease of $\eb$ with $\UAB$ with no local minimum. We illustrate this in Fig.~\ref{app:1d;fig:eb}, where we show the binding energy $\eb$ for an AABB system in a one-dimensional lattice with periodic boundary conditions. We employ Eq.~(\ref{sec:results;sub:eb;eq:ebAABB}) to calculate the binding energy.

Panel (a) shows that $\eb$ indeed simply decreases with $\UAB$. This can also be appreciated in panel (b), where the different curves always decrease with decreasing interspecies interaction. We have checked that this behavior appears for any lattice size in one dimension and for other numbers of particles.

Nevertheless,  $\eb$ in one dimension does show a local minimum if plotted as a function of $U$. This has been reported in previous works (see, for example, Ref.~\cite{morera_universal_2021}), and can be appreciated from Fig.~\ref{app:1d;fig:eb}(a) by following vertical lines at a constant $\UAB\lesssim 0.6$. However, this occurs for any dimensionality. 

\newpage 

\bibliography{biblio}

\begin{thebibliography}{74}%
\makeatletter
\providecommand \@ifxundefined [1]{%
 \@ifx{#1\undefined}
}%
\providecommand \@ifnum [1]{%
 \ifnum #1\expandafter \@firstoftwo
 \else \expandafter \@secondoftwo
 \fi
}%
\providecommand \@ifx [1]{%
 \ifx #1\expandafter \@firstoftwo
 \else \expandafter \@secondoftwo
 \fi
}%
\providecommand \natexlab [1]{#1}%
\providecommand \enquote  [1]{``#1''}%
\providecommand \bibnamefont  [1]{#1}%
\providecommand \bibfnamefont [1]{#1}%
\providecommand \citenamefont [1]{#1}%
\providecommand \href@noop [0]{\@secondoftwo}%
\providecommand \href [0]{\begingroup \@sanitize@url \@href}%
\providecommand \@href[1]{\@@startlink{#1}\@@href}%
\providecommand \@@href[1]{\endgroup#1\@@endlink}%
\providecommand \@sanitize@url [0]{\catcode `\\12\catcode `\$12\catcode
  `\&12\catcode `\#12\catcode `\^12\catcode `\_12\catcode `\%12\relax}%
\providecommand \@@startlink[1]{}%
\providecommand \@@endlink[0]{}%
\providecommand \url  [0]{\begingroup\@sanitize@url \@url }%
\providecommand \@url [1]{\endgroup\@href {#1}{\urlprefix }}%
\providecommand \urlprefix  [0]{URL }%
\providecommand \Eprint [0]{\href }%
\providecommand \doibase [0]{https://doi.org/}%
\providecommand \selectlanguage [0]{\@gobble}%
\providecommand \bibinfo  [0]{\@secondoftwo}%
\providecommand \bibfield  [0]{\@secondoftwo}%
\providecommand \translation [1]{[#1]}%
\providecommand \BibitemOpen [0]{}%
\providecommand \bibitemStop [0]{}%
\providecommand \bibitemNoStop [0]{.\EOS\space}%
\providecommand \EOS [0]{\spacefactor3000\relax}%
\providecommand \BibitemShut  [1]{\csname bibitem#1\endcsname}%
\let\auto@bib@innerbib\@empty
\bibitem [{\citenamefont {Sowiński}\ and\ \citenamefont
  {García-March}(2019)}]{sowinski_one-dimensional_2019}%
  \BibitemOpen
  \bibfield  {author} {\bibinfo {author} {\bibfnamefont {T.}~\bibnamefont
  {Sowiński}}\ and\ \bibinfo {author} {\bibfnamefont {M.~A.}\ \bibnamefont
  {García-March}},\ }\href {https://doi.org/10.1088/1361-6633/ab3a80}
  {\bibfield  {journal} {\bibinfo  {journal} {Reports on Progress in Physics}\
  }\textbf {\bibinfo {volume} {82}},\ \bibinfo {pages} {104401} (\bibinfo
  {year} {2019})}\BibitemShut {NoStop}%
\bibitem [{\citenamefont {Baroni}\ \emph {et~al.}(2024)\citenamefont {Baroni},
  \citenamefont {Lamporesi},\ and\ \citenamefont
  {Zaccanti}}]{baroni_quantum_2024}%
  \BibitemOpen
  \bibfield  {author} {\bibinfo {author} {\bibfnamefont {C.}~\bibnamefont
  {Baroni}}, \bibinfo {author} {\bibfnamefont {G.}~\bibnamefont {Lamporesi}},\
  and\ \bibinfo {author} {\bibfnamefont {M.}~\bibnamefont {Zaccanti}},\ }\href
  {https://doi.org/10.1038/s42254-024-00773-6} {\bibfield  {journal} {\bibinfo
  {journal} {Nature Reviews Physics}\ }\textbf {\bibinfo {volume} {6}},\
  \bibinfo {pages} {736} (\bibinfo {year} {2024})}\BibitemShut {NoStop}%
\bibitem [{\citenamefont {Fil}\ and\ \citenamefont
  {Shevchenko}(2005)}]{fil_nondissipative_2005}%
  \BibitemOpen
  \bibfield  {author} {\bibinfo {author} {\bibfnamefont {D.~V.}\ \bibnamefont
  {Fil}}\ and\ \bibinfo {author} {\bibfnamefont {S.~I.}\ \bibnamefont
  {Shevchenko}},\ }\href {https://doi.org/10.1103/PhysRevA.72.013616}
  {\bibfield  {journal} {\bibinfo  {journal} {Physical Review A}\ }\textbf
  {\bibinfo {volume} {72}},\ \bibinfo {pages} {013616} (\bibinfo {year}
  {2005})}\BibitemShut {NoStop}%
\bibitem [{\citenamefont {Nespolo}\ \emph {et~al.}(2017)\citenamefont
  {Nespolo}, \citenamefont {Astrakharchik},\ and\ \citenamefont
  {Recati}}]{nespolo_andreevbashkin_2017}%
  \BibitemOpen
  \bibfield  {author} {\bibinfo {author} {\bibfnamefont {J.}~\bibnamefont
  {Nespolo}}, \bibinfo {author} {\bibfnamefont {G.~E.}\ \bibnamefont
  {Astrakharchik}},\ and\ \bibinfo {author} {\bibfnamefont {A.}~\bibnamefont
  {Recati}},\ }\href {https://doi.org/10.1088/1367-2630/aa93a0} {\bibfield
  {journal} {\bibinfo  {journal} {New Journal of Physics}\ }\textbf {\bibinfo
  {volume} {19}},\ \bibinfo {pages} {125005} (\bibinfo {year}
  {2017})}\BibitemShut {NoStop}%
\bibitem [{\citenamefont {Juliá-Díaz}\ \emph {et~al.}(2009)\citenamefont
  {Juliá-Díaz}, \citenamefont {Guilleumas}, \citenamefont {Lewenstein},
  \citenamefont {Polls},\ and\ \citenamefont
  {Sanpera}}]{julia-diaz_josephson_2009}%
  \BibitemOpen
  \bibfield  {author} {\bibinfo {author} {\bibfnamefont {B.}~\bibnamefont
  {Juliá-Díaz}}, \bibinfo {author} {\bibfnamefont {M.}~\bibnamefont
  {Guilleumas}}, \bibinfo {author} {\bibfnamefont {M.}~\bibnamefont
  {Lewenstein}}, \bibinfo {author} {\bibfnamefont {A.}~\bibnamefont {Polls}},\
  and\ \bibinfo {author} {\bibfnamefont {A.}~\bibnamefont {Sanpera}},\ }\href
  {https://doi.org/10.1103/PhysRevA.80.023616} {\bibfield  {journal} {\bibinfo
  {journal} {Physical Review A}\ }\textbf {\bibinfo {volume} {80}},\ \bibinfo
  {pages} {023616} (\bibinfo {year} {2009})}\BibitemShut {NoStop}%
\bibitem [{\citenamefont {Melé-Messeguer}\ \emph {et~al.}(2011)\citenamefont
  {Melé-Messeguer}, \citenamefont {Juliá-Díaz}, \citenamefont {Guilleumas},
  \citenamefont {Polls},\ and\ \citenamefont
  {Sanpera}}]{mele-messeguer_weakly_2011}%
  \BibitemOpen
  \bibfield  {author} {\bibinfo {author} {\bibfnamefont {M.}~\bibnamefont
  {Melé-Messeguer}}, \bibinfo {author} {\bibfnamefont {B.}~\bibnamefont
  {Juliá-Díaz}}, \bibinfo {author} {\bibfnamefont {M.}~\bibnamefont
  {Guilleumas}}, \bibinfo {author} {\bibfnamefont {A.}~\bibnamefont {Polls}},\
  and\ \bibinfo {author} {\bibfnamefont {A.}~\bibnamefont {Sanpera}},\ }\href
  {https://doi.org/10.1088/1367-2630/13/3/033012} {\bibfield  {journal}
  {\bibinfo  {journal} {New Journal of Physics}\ }\textbf {\bibinfo {volume}
  {13}},\ \bibinfo {pages} {033012} (\bibinfo {year} {2011})}\BibitemShut
  {NoStop}%
\bibitem [{\citenamefont {Morales-Molina}\ \emph {et~al.}(2020)\citenamefont
  {Morales-Molina}, \citenamefont {Reyes},\ and\ \citenamefont
  {Arévalo}}]{morales-molina_harnessing_2020}%
  \BibitemOpen
  \bibfield  {author} {\bibinfo {author} {\bibfnamefont {L.}~\bibnamefont
  {Morales-Molina}}, \bibinfo {author} {\bibfnamefont {S.~A.}\ \bibnamefont
  {Reyes}},\ and\ \bibinfo {author} {\bibfnamefont {E.}~\bibnamefont
  {Arévalo}},\ }\href {https://doi.org/10.1209/0295-5075/131/36001} {\bibfield
   {journal} {\bibinfo  {journal} {EPL (Europhysics Letters)}\ }\textbf
  {\bibinfo {volume} {131}},\ \bibinfo {pages} {36001} (\bibinfo {year}
  {2020})}\BibitemShut {NoStop}%
\bibitem [{\citenamefont {Spehner}\ \emph {et~al.}(2021)\citenamefont
  {Spehner}, \citenamefont {Morales-Molina},\ and\ \citenamefont
  {Reyes}}]{spehner_persistent_2021}%
  \BibitemOpen
  \bibfield  {author} {\bibinfo {author} {\bibfnamefont {D.}~\bibnamefont
  {Spehner}}, \bibinfo {author} {\bibfnamefont {L.}~\bibnamefont
  {Morales-Molina}},\ and\ \bibinfo {author} {\bibfnamefont {S.~A.}\
  \bibnamefont {Reyes}},\ }\href {https://doi.org/10.1088/1367-2630/abeebb}
  {\bibfield  {journal} {\bibinfo  {journal} {New Journal of Physics}\ }\textbf
  {\bibinfo {volume} {23}},\ \bibinfo {pages} {063025} (\bibinfo {year}
  {2021})}\BibitemShut {NoStop}%
\bibitem [{\citenamefont {Guijarro}\ \emph {et~al.}(2018)\citenamefont
  {Guijarro}, \citenamefont {Pricoupenko}, \citenamefont {Astrakharchik},
  \citenamefont {Boronat},\ and\ \citenamefont
  {Petrov}}]{guijarro_one-dimensional_2018}%
  \BibitemOpen
  \bibfield  {author} {\bibinfo {author} {\bibfnamefont {G.}~\bibnamefont
  {Guijarro}}, \bibinfo {author} {\bibfnamefont {A.}~\bibnamefont
  {Pricoupenko}}, \bibinfo {author} {\bibfnamefont {G.~E.}\ \bibnamefont
  {Astrakharchik}}, \bibinfo {author} {\bibfnamefont {J.}~\bibnamefont
  {Boronat}},\ and\ \bibinfo {author} {\bibfnamefont {D.~S.}\ \bibnamefont
  {Petrov}},\ }\href {https://doi.org/10.1103/PhysRevA.97.061605} {\bibfield
  {journal} {\bibinfo  {journal} {Physical Review A}\ }\textbf {\bibinfo
  {volume} {97}},\ \bibinfo {pages} {061605} (\bibinfo {year}
  {2018})}\BibitemShut {NoStop}%
\bibitem [{\citenamefont {Guijarro}\ \emph {et~al.}(2020)\citenamefont
  {Guijarro}, \citenamefont {Astrakharchik}, \citenamefont {Boronat},
  \citenamefont {Bazak},\ and\ \citenamefont
  {Petrov}}]{guijarro_few-body_2020}%
  \BibitemOpen
  \bibfield  {author} {\bibinfo {author} {\bibfnamefont {G.}~\bibnamefont
  {Guijarro}}, \bibinfo {author} {\bibfnamefont {G.~E.}\ \bibnamefont
  {Astrakharchik}}, \bibinfo {author} {\bibfnamefont {J.}~\bibnamefont
  {Boronat}}, \bibinfo {author} {\bibfnamefont {B.}~\bibnamefont {Bazak}},\
  and\ \bibinfo {author} {\bibfnamefont {D.~S.}\ \bibnamefont {Petrov}},\
  }\href {https://doi.org/10.1103/PhysRevA.101.041602} {\bibfield  {journal}
  {\bibinfo  {journal} {Physical Review A}\ }\textbf {\bibinfo {volume}
  {101}},\ \bibinfo {pages} {041602} (\bibinfo {year} {2020})}\BibitemShut
  {NoStop}%
\bibitem [{\citenamefont {Liu}\ \emph {et~al.}(2021)\citenamefont {Liu},
  \citenamefont {Yu},\ and\ \citenamefont {Chen}}]{liu_three-body_2021}%
  \BibitemOpen
  \bibfield  {author} {\bibinfo {author} {\bibfnamefont {Y.}~\bibnamefont
  {Liu}}, \bibinfo {author} {\bibfnamefont {Y.-C.}\ \bibnamefont {Yu}},\ and\
  \bibinfo {author} {\bibfnamefont {S.}~\bibnamefont {Chen}},\ }\href
  {https://doi.org/10.1103/PhysRevA.104.033303} {\bibfield  {journal} {\bibinfo
   {journal} {Physical Review A}\ }\textbf {\bibinfo {volume} {104}},\ \bibinfo
  {pages} {033303} (\bibinfo {year} {2021})}\BibitemShut {NoStop}%
\bibitem [{\citenamefont {Petrov}(2015)}]{petrov_quantum_2015}%
  \BibitemOpen
  \bibfield  {author} {\bibinfo {author} {\bibfnamefont {D.}~\bibnamefont
  {Petrov}},\ }\href {https://doi.org/10.1103/PhysRevLett.115.155302}
  {\bibfield  {journal} {\bibinfo  {journal} {Physical Review Letters}\
  }\textbf {\bibinfo {volume} {115}},\ \bibinfo {pages} {155302} (\bibinfo
  {year} {2015})}\BibitemShut {NoStop}%
\bibitem [{\citenamefont {Petrov}\ and\ \citenamefont
  {Astrakharchik}(2016)}]{petrov_ultradilute_2016}%
  \BibitemOpen
  \bibfield  {author} {\bibinfo {author} {\bibfnamefont {D.}~\bibnamefont
  {Petrov}}\ and\ \bibinfo {author} {\bibfnamefont {G.}~\bibnamefont
  {Astrakharchik}},\ }\href {https://doi.org/10.1103/PhysRevLett.117.100401}
  {\bibfield  {journal} {\bibinfo  {journal} {Physical Review Letters}\
  }\textbf {\bibinfo {volume} {117}},\ \bibinfo {pages} {100401} (\bibinfo
  {year} {2016})}\BibitemShut {NoStop}%
\bibitem [{\citenamefont {Cabrera}\ \emph {et~al.}(2018)\citenamefont
  {Cabrera}, \citenamefont {Tanzi}, \citenamefont {Sanz}, \citenamefont
  {Naylor}, \citenamefont {Thomas}, \citenamefont {Cheiney},\ and\
  \citenamefont {Tarruell}}]{cabrera_quantum_2018}%
  \BibitemOpen
  \bibfield  {author} {\bibinfo {author} {\bibfnamefont {C.~R.}\ \bibnamefont
  {Cabrera}}, \bibinfo {author} {\bibfnamefont {L.}~\bibnamefont {Tanzi}},
  \bibinfo {author} {\bibfnamefont {J.}~\bibnamefont {Sanz}}, \bibinfo {author}
  {\bibfnamefont {B.}~\bibnamefont {Naylor}}, \bibinfo {author} {\bibfnamefont
  {P.}~\bibnamefont {Thomas}}, \bibinfo {author} {\bibfnamefont
  {P.}~\bibnamefont {Cheiney}},\ and\ \bibinfo {author} {\bibfnamefont
  {L.}~\bibnamefont {Tarruell}},\ }\href
  {https://doi.org/10.1126/science.aao5686} {\bibfield  {journal} {\bibinfo
  {journal} {Science}\ }\textbf {\bibinfo {volume} {359}},\ \bibinfo {pages}
  {301} (\bibinfo {year} {2018})}\BibitemShut {NoStop}%
\bibitem [{\citenamefont {Semeghini}\ \emph {et~al.}(2018)\citenamefont
  {Semeghini}, \citenamefont {Ferioli}, \citenamefont {Masi}, \citenamefont
  {Mazzinghi}, \citenamefont {Wolswijk}, \citenamefont {Minardi}, \citenamefont
  {Modugno}, \citenamefont {Modugno}, \citenamefont {Inguscio},\ and\
  \citenamefont {Fattori}}]{semeghini_self-bound_2018}%
  \BibitemOpen
  \bibfield  {author} {\bibinfo {author} {\bibfnamefont {G.}~\bibnamefont
  {Semeghini}}, \bibinfo {author} {\bibfnamefont {G.}~\bibnamefont {Ferioli}},
  \bibinfo {author} {\bibfnamefont {L.}~\bibnamefont {Masi}}, \bibinfo {author}
  {\bibfnamefont {C.}~\bibnamefont {Mazzinghi}}, \bibinfo {author}
  {\bibfnamefont {L.}~\bibnamefont {Wolswijk}}, \bibinfo {author}
  {\bibfnamefont {F.}~\bibnamefont {Minardi}}, \bibinfo {author} {\bibfnamefont
  {M.}~\bibnamefont {Modugno}}, \bibinfo {author} {\bibfnamefont
  {G.}~\bibnamefont {Modugno}}, \bibinfo {author} {\bibfnamefont
  {M.}~\bibnamefont {Inguscio}},\ and\ \bibinfo {author} {\bibfnamefont
  {M.}~\bibnamefont {Fattori}},\ }\href
  {https://doi.org/10.1103/PhysRevLett.120.235301} {\bibfield  {journal}
  {\bibinfo  {journal} {Physical Review Letters}\ }\textbf {\bibinfo {volume}
  {120}},\ \bibinfo {pages} {235301} (\bibinfo {year} {2018})}\BibitemShut
  {NoStop}%
\bibitem [{\citenamefont {Cheiney}\ \emph {et~al.}(2018)\citenamefont
  {Cheiney}, \citenamefont {Cabrera}, \citenamefont {Sanz}, \citenamefont
  {Naylor}, \citenamefont {Tanzi},\ and\ \citenamefont
  {Tarruell}}]{cheiney_bright_2018}%
  \BibitemOpen
  \bibfield  {author} {\bibinfo {author} {\bibfnamefont {P.}~\bibnamefont
  {Cheiney}}, \bibinfo {author} {\bibfnamefont {C.}~\bibnamefont {Cabrera}},
  \bibinfo {author} {\bibfnamefont {J.}~\bibnamefont {Sanz}}, \bibinfo {author}
  {\bibfnamefont {B.}~\bibnamefont {Naylor}}, \bibinfo {author} {\bibfnamefont
  {L.}~\bibnamefont {Tanzi}},\ and\ \bibinfo {author} {\bibfnamefont
  {L.}~\bibnamefont {Tarruell}},\ }\href
  {https://doi.org/10.1103/PhysRevLett.120.135301} {\bibfield  {journal}
  {\bibinfo  {journal} {Physical Review Letters}\ }\textbf {\bibinfo {volume}
  {120}},\ \bibinfo {pages} {135301} (\bibinfo {year} {2018})}\BibitemShut
  {NoStop}%
\bibitem [{\citenamefont {Ferioli}\ \emph {et~al.}(2019)\citenamefont
  {Ferioli}, \citenamefont {Semeghini}, \citenamefont {Masi}, \citenamefont
  {Giusti}, \citenamefont {Modugno}, \citenamefont {Inguscio}, \citenamefont
  {Gallemí}, \citenamefont {Recati},\ and\ \citenamefont
  {Fattori}}]{ferioli_collisions_2019}%
  \BibitemOpen
  \bibfield  {author} {\bibinfo {author} {\bibfnamefont {G.}~\bibnamefont
  {Ferioli}}, \bibinfo {author} {\bibfnamefont {G.}~\bibnamefont {Semeghini}},
  \bibinfo {author} {\bibfnamefont {L.}~\bibnamefont {Masi}}, \bibinfo {author}
  {\bibfnamefont {G.}~\bibnamefont {Giusti}}, \bibinfo {author} {\bibfnamefont
  {G.}~\bibnamefont {Modugno}}, \bibinfo {author} {\bibfnamefont
  {M.}~\bibnamefont {Inguscio}}, \bibinfo {author} {\bibfnamefont
  {A.}~\bibnamefont {Gallemí}}, \bibinfo {author} {\bibfnamefont
  {A.}~\bibnamefont {Recati}},\ and\ \bibinfo {author} {\bibfnamefont
  {M.}~\bibnamefont {Fattori}},\ }\href
  {https://doi.org/10.1103/PhysRevLett.122.090401} {\bibfield  {journal}
  {\bibinfo  {journal} {Physical Review Letters}\ }\textbf {\bibinfo {volume}
  {122}},\ \bibinfo {pages} {090401} (\bibinfo {year} {2019})}\BibitemShut
  {NoStop}%
\bibitem [{\citenamefont {D'Errico}\ \emph {et~al.}(2019)\citenamefont
  {D'Errico}, \citenamefont {Burchianti}, \citenamefont {Prevedelli},
  \citenamefont {Salasnich}, \citenamefont {Ancilotto}, \citenamefont
  {Modugno}, \citenamefont {Minardi},\ and\ \citenamefont
  {Fort}}]{derrico_observation_2019}%
  \BibitemOpen
  \bibfield  {author} {\bibinfo {author} {\bibfnamefont {C.}~\bibnamefont
  {D'Errico}}, \bibinfo {author} {\bibfnamefont {A.}~\bibnamefont
  {Burchianti}}, \bibinfo {author} {\bibfnamefont {M.}~\bibnamefont
  {Prevedelli}}, \bibinfo {author} {\bibfnamefont {L.}~\bibnamefont
  {Salasnich}}, \bibinfo {author} {\bibfnamefont {F.}~\bibnamefont
  {Ancilotto}}, \bibinfo {author} {\bibfnamefont {M.}~\bibnamefont {Modugno}},
  \bibinfo {author} {\bibfnamefont {F.}~\bibnamefont {Minardi}},\ and\ \bibinfo
  {author} {\bibfnamefont {C.}~\bibnamefont {Fort}},\ }\href
  {https://doi.org/10.1103/PhysRevResearch.1.033155} {\bibfield  {journal}
  {\bibinfo  {journal} {Physical Review Research}\ }\textbf {\bibinfo {volume}
  {1}},\ \bibinfo {pages} {033155} (\bibinfo {year} {2019})}\BibitemShut
  {NoStop}%
\bibitem [{\citenamefont {Hu}\ and\ \citenamefont
  {Liu}(2020)}]{hu_consistent_2020}%
  \BibitemOpen
  \bibfield  {author} {\bibinfo {author} {\bibfnamefont {H.}~\bibnamefont
  {Hu}}\ and\ \bibinfo {author} {\bibfnamefont {X.-J.}\ \bibnamefont {Liu}},\
  }\href {https://doi.org/10.1103/PhysRevLett.125.195302} {\bibfield  {journal}
  {\bibinfo  {journal} {Physical Review Letters}\ }\textbf {\bibinfo {volume}
  {125}},\ \bibinfo {pages} {195302} (\bibinfo {year} {2020})}\BibitemShut
  {NoStop}%
\bibitem [{\citenamefont {Hu}\ \emph {et~al.}(2020)\citenamefont {Hu},
  \citenamefont {Wang},\ and\ \citenamefont {Liu}}]{hu_microscopic_2020-1}%
  \BibitemOpen
  \bibfield  {author} {\bibinfo {author} {\bibfnamefont {H.}~\bibnamefont
  {Hu}}, \bibinfo {author} {\bibfnamefont {J.}~\bibnamefont {Wang}},\ and\
  \bibinfo {author} {\bibfnamefont {X.-J.}\ \bibnamefont {Liu}},\ }\href
  {https://doi.org/10.1103/PhysRevA.102.043301} {\bibfield  {journal} {\bibinfo
   {journal} {Physical Review A}\ }\textbf {\bibinfo {volume} {102}},\ \bibinfo
  {pages} {043301} (\bibinfo {year} {2020})}\BibitemShut {NoStop}%
\bibitem [{\citenamefont {Ota}\ and\ \citenamefont
  {Astrakharchik}(2020)}]{ota_beyond_2020}%
  \BibitemOpen
  \bibfield  {author} {\bibinfo {author} {\bibfnamefont {M.}~\bibnamefont
  {Ota}}\ and\ \bibinfo {author} {\bibfnamefont {G.}~\bibnamefont
  {Astrakharchik}},\ }\href {https://doi.org/10.21468/SciPostPhys.9.2.020}
  {\bibfield  {journal} {\bibinfo  {journal} {SciPost Physics}\ }\textbf
  {\bibinfo {volume} {9}},\ \bibinfo {pages} {020} (\bibinfo {year}
  {2020})}\BibitemShut {NoStop}%
\bibitem [{\citenamefont {Tylutki}\ \emph {et~al.}(2020)\citenamefont
  {Tylutki}, \citenamefont {Astrakharchik}, \citenamefont {Malomed},\ and\
  \citenamefont {Petrov}}]{tylutki_collective_2020}%
  \BibitemOpen
  \bibfield  {author} {\bibinfo {author} {\bibfnamefont {M.}~\bibnamefont
  {Tylutki}}, \bibinfo {author} {\bibfnamefont {G.~E.}\ \bibnamefont
  {Astrakharchik}}, \bibinfo {author} {\bibfnamefont {B.~A.}\ \bibnamefont
  {Malomed}},\ and\ \bibinfo {author} {\bibfnamefont {D.~S.}\ \bibnamefont
  {Petrov}},\ }\href {https://doi.org/10.1103/PhysRevA.101.051601} {\bibfield
  {journal} {\bibinfo  {journal} {Physical Review A}\ }\textbf {\bibinfo
  {volume} {101}},\ \bibinfo {pages} {051601} (\bibinfo {year}
  {2020})}\BibitemShut {NoStop}%
\bibitem [{\citenamefont {Guebli}\ and\ \citenamefont
  {Boudjemâa}(2021)}]{guebli_quantum_2021}%
  \BibitemOpen
  \bibfield  {author} {\bibinfo {author} {\bibfnamefont {N.}~\bibnamefont
  {Guebli}}\ and\ \bibinfo {author} {\bibfnamefont {A.}~\bibnamefont
  {Boudjemâa}},\ }\href {https://doi.org/10.1103/PhysRevA.104.023310}
  {\bibfield  {journal} {\bibinfo  {journal} {Physical Review A}\ }\textbf
  {\bibinfo {volume} {104}},\ \bibinfo {pages} {023310} (\bibinfo {year}
  {2021})}\BibitemShut {NoStop}%
\bibitem [{\citenamefont {Xiong}\ and\ \citenamefont
  {Yin}(2022)}]{xiong_effective_2022}%
  \BibitemOpen
  \bibfield  {author} {\bibinfo {author} {\bibfnamefont {Y.}~\bibnamefont
  {Xiong}}\ and\ \bibinfo {author} {\bibfnamefont {L.}~\bibnamefont {Yin}},\
  }\href {https://doi.org/10.1103/PhysRevA.105.053305} {\bibfield  {journal}
  {\bibinfo  {journal} {Physical Review A}\ }\textbf {\bibinfo {volume}
  {105}},\ \bibinfo {pages} {053305} (\bibinfo {year} {2022})}\BibitemShut
  {NoStop}%
\bibitem [{\citenamefont {Pan}\ \emph {et~al.}(2022)\citenamefont {Pan},
  \citenamefont {Yi},\ and\ \citenamefont {Shi}}]{pan_quantum_2022}%
  \BibitemOpen
  \bibfield  {author} {\bibinfo {author} {\bibfnamefont {J.}~\bibnamefont
  {Pan}}, \bibinfo {author} {\bibfnamefont {S.}~\bibnamefont {Yi}},\ and\
  \bibinfo {author} {\bibfnamefont {T.}~\bibnamefont {Shi}},\ }\href
  {https://doi.org/10.1103/PhysRevResearch.4.043018} {\bibfield  {journal}
  {\bibinfo  {journal} {Physical Review Research}\ }\textbf {\bibinfo {volume}
  {4}},\ \bibinfo {pages} {043018} (\bibinfo {year} {2022})}\BibitemShut
  {NoStop}%
\bibitem [{\citenamefont {Boudjemâa}\ and\ \citenamefont
  {Abbas}(2023)}]{boudjemaa_quantum_2023}%
  \BibitemOpen
  \bibfield  {author} {\bibinfo {author} {\bibfnamefont {A.}~\bibnamefont
  {Boudjemâa}}\ and\ \bibinfo {author} {\bibfnamefont {K.}~\bibnamefont
  {Abbas}},\ }\href {https://doi.org/10.1088/1367-2630/acf8ed} {\bibfield
  {journal} {\bibinfo  {journal} {New Journal of Physics}\ }\textbf {\bibinfo
  {volume} {25}},\ \bibinfo {pages} {093052} (\bibinfo {year}
  {2023})}\BibitemShut {NoStop}%
\bibitem [{\citenamefont {Spada}\ \emph {et~al.}(2024)\citenamefont {Spada},
  \citenamefont {Pilati},\ and\ \citenamefont {Giorgini}}]{spada_quantum_2024}%
  \BibitemOpen
  \bibfield  {author} {\bibinfo {author} {\bibfnamefont {G.}~\bibnamefont
  {Spada}}, \bibinfo {author} {\bibfnamefont {S.}~\bibnamefont {Pilati}},\ and\
  \bibinfo {author} {\bibfnamefont {S.}~\bibnamefont {Giorgini}},\ }\href
  {https://doi.org/10.1103/PhysRevLett.133.083401} {\bibfield  {journal}
  {\bibinfo  {journal} {Physical Review Letters}\ }\textbf {\bibinfo {volume}
  {133}},\ \bibinfo {pages} {083401} (\bibinfo {year} {2024})}\BibitemShut
  {NoStop}%
\bibitem [{\citenamefont {Lahaye}\ \emph {et~al.}(2009)\citenamefont {Lahaye},
  \citenamefont {Menotti}, \citenamefont {Santos}, \citenamefont {Lewenstein},\
  and\ \citenamefont {Pfau}}]{lahaye_physics_2009}%
  \BibitemOpen
  \bibfield  {author} {\bibinfo {author} {\bibfnamefont {T.}~\bibnamefont
  {Lahaye}}, \bibinfo {author} {\bibfnamefont {C.}~\bibnamefont {Menotti}},
  \bibinfo {author} {\bibfnamefont {L.}~\bibnamefont {Santos}}, \bibinfo
  {author} {\bibfnamefont {M.}~\bibnamefont {Lewenstein}},\ and\ \bibinfo
  {author} {\bibfnamefont {T.}~\bibnamefont {Pfau}},\ }\href
  {https://doi.org/10.1088/0034-4885/72/12/126401} {\bibfield  {journal}
  {\bibinfo  {journal} {Reports on Progress in Physics}\ }\textbf {\bibinfo
  {volume} {72}},\ \bibinfo {pages} {126401} (\bibinfo {year}
  {2009})}\BibitemShut {NoStop}%
\bibitem [{\citenamefont {Baranov}\ \emph {et~al.}(2012)\citenamefont
  {Baranov}, \citenamefont {Dalmonte}, \citenamefont {Pupillo},\ and\
  \citenamefont {Zoller}}]{baranov_condensed_2012}%
  \BibitemOpen
  \bibfield  {author} {\bibinfo {author} {\bibfnamefont {M.~A.}\ \bibnamefont
  {Baranov}}, \bibinfo {author} {\bibfnamefont {M.}~\bibnamefont {Dalmonte}},
  \bibinfo {author} {\bibfnamefont {G.}~\bibnamefont {Pupillo}},\ and\ \bibinfo
  {author} {\bibfnamefont {P.}~\bibnamefont {Zoller}},\ }\href
  {https://doi.org/10.1021/cr2003568} {\bibfield  {journal} {\bibinfo
  {journal} {Chemical Reviews}\ }\textbf {\bibinfo {volume} {112}},\ \bibinfo
  {pages} {5012} (\bibinfo {year} {2012})}\BibitemShut {NoStop}%
\bibitem [{\citenamefont {Trefzger}\ \emph {et~al.}(2011)\citenamefont
  {Trefzger}, \citenamefont {Menotti}, \citenamefont {Capogrosso-Sansone},\
  and\ \citenamefont {Lewenstein}}]{trefzger_ultracold_2011}%
  \BibitemOpen
  \bibfield  {author} {\bibinfo {author} {\bibfnamefont {C.}~\bibnamefont
  {Trefzger}}, \bibinfo {author} {\bibfnamefont {C.}~\bibnamefont {Menotti}},
  \bibinfo {author} {\bibfnamefont {B.}~\bibnamefont {Capogrosso-Sansone}},\
  and\ \bibinfo {author} {\bibfnamefont {M.}~\bibnamefont {Lewenstein}},\
  }\href {https://doi.org/10.1088/0953-4075/44/19/193001} {\bibfield  {journal}
  {\bibinfo  {journal} {Journal of Physics B: Atomic, Molecular and Optical
  Physics}\ }\textbf {\bibinfo {volume} {44}},\ \bibinfo {pages} {193001}
  (\bibinfo {year} {2011})}\BibitemShut {NoStop}%
\bibitem [{\citenamefont {Chomaz}\ \emph {et~al.}(2023)\citenamefont {Chomaz},
  \citenamefont {Ferrier-Barbut}, \citenamefont {Ferlaino}, \citenamefont
  {Laburthe-Tolra}, \citenamefont {Lev},\ and\ \citenamefont
  {Pfau}}]{chomaz_dipolar_2023}%
  \BibitemOpen
  \bibfield  {author} {\bibinfo {author} {\bibfnamefont {L.}~\bibnamefont
  {Chomaz}}, \bibinfo {author} {\bibfnamefont {I.}~\bibnamefont
  {Ferrier-Barbut}}, \bibinfo {author} {\bibfnamefont {F.}~\bibnamefont
  {Ferlaino}}, \bibinfo {author} {\bibfnamefont {B.}~\bibnamefont
  {Laburthe-Tolra}}, \bibinfo {author} {\bibfnamefont {B.~L.}\ \bibnamefont
  {Lev}},\ and\ \bibinfo {author} {\bibfnamefont {T.}~\bibnamefont {Pfau}},\
  }\href {https://doi.org/10.1088/1361-6633/aca814} {\bibfield  {journal}
  {\bibinfo  {journal} {Reports on Progress in Physics}\ }\textbf {\bibinfo
  {volume} {86}},\ \bibinfo {pages} {026401} (\bibinfo {year}
  {2023})}\BibitemShut {NoStop}%
\bibitem [{\citenamefont {Schmitt}\ \emph {et~al.}(2016)\citenamefont
  {Schmitt}, \citenamefont {Wenzel}, \citenamefont {Böttcher}, \citenamefont
  {Ferrier-Barbut},\ and\ \citenamefont {Pfau}}]{schmitt_self-bound_2016}%
  \BibitemOpen
  \bibfield  {author} {\bibinfo {author} {\bibfnamefont {M.}~\bibnamefont
  {Schmitt}}, \bibinfo {author} {\bibfnamefont {M.}~\bibnamefont {Wenzel}},
  \bibinfo {author} {\bibfnamefont {F.}~\bibnamefont {Böttcher}}, \bibinfo
  {author} {\bibfnamefont {I.}~\bibnamefont {Ferrier-Barbut}},\ and\ \bibinfo
  {author} {\bibfnamefont {T.}~\bibnamefont {Pfau}},\ }\href
  {https://doi.org/10.1038/nature20126} {\bibfield  {journal} {\bibinfo
  {journal} {Nature}\ }\textbf {\bibinfo {volume} {539}},\ \bibinfo {pages}
  {259} (\bibinfo {year} {2016})}\BibitemShut {NoStop}%
\bibitem [{\citenamefont {Ferrier-Barbut}\ \emph {et~al.}(2016)\citenamefont
  {Ferrier-Barbut}, \citenamefont {Kadau}, \citenamefont {Schmitt},
  \citenamefont {Wenzel},\ and\ \citenamefont
  {Pfau}}]{ferrier-barbut_observation_2016}%
  \BibitemOpen
  \bibfield  {author} {\bibinfo {author} {\bibfnamefont {I.}~\bibnamefont
  {Ferrier-Barbut}}, \bibinfo {author} {\bibfnamefont {H.}~\bibnamefont
  {Kadau}}, \bibinfo {author} {\bibfnamefont {M.}~\bibnamefont {Schmitt}},
  \bibinfo {author} {\bibfnamefont {M.}~\bibnamefont {Wenzel}},\ and\ \bibinfo
  {author} {\bibfnamefont {T.}~\bibnamefont {Pfau}},\ }\href
  {https://doi.org/10.1103/PhysRevLett.116.215301} {\bibfield  {journal}
  {\bibinfo  {journal} {Physical Review Letters}\ }\textbf {\bibinfo {volume}
  {116}},\ \bibinfo {pages} {215301} (\bibinfo {year} {2016})}\BibitemShut
  {NoStop}%
\bibitem [{\citenamefont {Chomaz}\ \emph {et~al.}(2016)\citenamefont {Chomaz},
  \citenamefont {Baier}, \citenamefont {Petter}, \citenamefont {Mark},
  \citenamefont {Wächtler}, \citenamefont {Santos},\ and\ \citenamefont
  {Ferlaino}}]{chomaz_quantum-fluctuation-driven_2016}%
  \BibitemOpen
  \bibfield  {author} {\bibinfo {author} {\bibfnamefont {L.}~\bibnamefont
  {Chomaz}}, \bibinfo {author} {\bibfnamefont {S.}~\bibnamefont {Baier}},
  \bibinfo {author} {\bibfnamefont {D.}~\bibnamefont {Petter}}, \bibinfo
  {author} {\bibfnamefont {M.}~\bibnamefont {Mark}}, \bibinfo {author}
  {\bibfnamefont {F.}~\bibnamefont {Wächtler}}, \bibinfo {author}
  {\bibfnamefont {L.}~\bibnamefont {Santos}},\ and\ \bibinfo {author}
  {\bibfnamefont {F.}~\bibnamefont {Ferlaino}},\ }\href
  {https://doi.org/10.1103/PhysRevX.6.041039} {\bibfield  {journal} {\bibinfo
  {journal} {Physical Review X}\ }\textbf {\bibinfo {volume} {6}},\ \bibinfo
  {pages} {041039} (\bibinfo {year} {2016})}\BibitemShut {NoStop}%
\bibitem [{\citenamefont {Klawunn}\ \emph {et~al.}(2010)\citenamefont
  {Klawunn}, \citenamefont {Pikovski},\ and\ \citenamefont
  {Santos}}]{klawunn_two-dimensional_2010}%
  \BibitemOpen
  \bibfield  {author} {\bibinfo {author} {\bibfnamefont {M.}~\bibnamefont
  {Klawunn}}, \bibinfo {author} {\bibfnamefont {A.}~\bibnamefont {Pikovski}},\
  and\ \bibinfo {author} {\bibfnamefont {L.}~\bibnamefont {Santos}},\ }\href
  {https://doi.org/10.1103/PhysRevA.82.044701} {\bibfield  {journal} {\bibinfo
  {journal} {Physical Review A}\ }\textbf {\bibinfo {volume} {82}},\ \bibinfo
  {pages} {044701} (\bibinfo {year} {2010})}\BibitemShut {NoStop}%
\bibitem [{\citenamefont {Wunsch}\ \emph {et~al.}(2011)\citenamefont {Wunsch},
  \citenamefont {Zinner}, \citenamefont {Mekhov}, \citenamefont {Huang},
  \citenamefont {Wang},\ and\ \citenamefont {Demler}}]{wunsch_few-body_2011}%
  \BibitemOpen
  \bibfield  {author} {\bibinfo {author} {\bibfnamefont {B.}~\bibnamefont
  {Wunsch}}, \bibinfo {author} {\bibfnamefont {N.~T.}\ \bibnamefont {Zinner}},
  \bibinfo {author} {\bibfnamefont {I.~B.}\ \bibnamefont {Mekhov}}, \bibinfo
  {author} {\bibfnamefont {S.-J.}\ \bibnamefont {Huang}}, \bibinfo {author}
  {\bibfnamefont {D.-W.}\ \bibnamefont {Wang}},\ and\ \bibinfo {author}
  {\bibfnamefont {E.}~\bibnamefont {Demler}},\ }\href
  {https://doi.org/10.1103/PhysRevLett.107.073201} {\bibfield  {journal}
  {\bibinfo  {journal} {Physical Review Letters}\ }\textbf {\bibinfo {volume}
  {107}},\ \bibinfo {pages} {073201} (\bibinfo {year} {2011})}\BibitemShut
  {NoStop}%
\bibitem [{\citenamefont {Volosniev}\ \emph {et~al.}(2012)\citenamefont
  {Volosniev}, \citenamefont {Fedorov}, \citenamefont {Jensen},\ and\
  \citenamefont {Zinner}}]{volosniev_few-body_2012}%
  \BibitemOpen
  \bibfield  {author} {\bibinfo {author} {\bibfnamefont {A.~G.}\ \bibnamefont
  {Volosniev}}, \bibinfo {author} {\bibfnamefont {D.~V.}\ \bibnamefont
  {Fedorov}}, \bibinfo {author} {\bibfnamefont {A.~S.}\ \bibnamefont
  {Jensen}},\ and\ \bibinfo {author} {\bibfnamefont {N.~T.}\ \bibnamefont
  {Zinner}},\ }\href {https://doi.org/10.1103/PhysRevA.85.023609} {\bibfield
  {journal} {\bibinfo  {journal} {Physical Review A}\ }\textbf {\bibinfo
  {volume} {85}},\ \bibinfo {pages} {023609} (\bibinfo {year}
  {2012})}\BibitemShut {NoStop}%
\bibitem [{\citenamefont {Volosniev}\ \emph {et~al.}(2013)\citenamefont
  {Volosniev}, \citenamefont {Armstrong}, \citenamefont {Fedorov},
  \citenamefont {Jensen}, \citenamefont {Valiente},\ and\ \citenamefont
  {Zinner}}]{volosniev_bound_2013}%
  \BibitemOpen
  \bibfield  {author} {\bibinfo {author} {\bibfnamefont {A.~G.}\ \bibnamefont
  {Volosniev}}, \bibinfo {author} {\bibfnamefont {J.~R.}\ \bibnamefont
  {Armstrong}}, \bibinfo {author} {\bibfnamefont {D.~V.}\ \bibnamefont
  {Fedorov}}, \bibinfo {author} {\bibfnamefont {A.~S.}\ \bibnamefont {Jensen}},
  \bibinfo {author} {\bibfnamefont {M.}~\bibnamefont {Valiente}},\ and\
  \bibinfo {author} {\bibfnamefont {N.~T.}\ \bibnamefont {Zinner}},\ }\href
  {https://doi.org/10.1088/1367-2630/15/4/043046} {\bibfield  {journal}
  {\bibinfo  {journal} {New Journal of Physics}\ }\textbf {\bibinfo {volume}
  {15}},\ \bibinfo {pages} {043046} (\bibinfo {year} {2013})}\BibitemShut
  {NoStop}%
\bibitem [{\citenamefont {Bloch}(2005)}]{bloch_ultracold_2005}%
  \BibitemOpen
  \bibfield  {author} {\bibinfo {author} {\bibfnamefont {I.}~\bibnamefont
  {Bloch}},\ }\href {https://doi.org/10.1038/nphys138} {\bibfield  {journal}
  {\bibinfo  {journal} {Nature Physics}\ }\textbf {\bibinfo {volume} {1}},\
  \bibinfo {pages} {23} (\bibinfo {year} {2005})}\BibitemShut {NoStop}%
\bibitem [{\citenamefont {Lewenstein}\ \emph {et~al.}(2012)\citenamefont
  {Lewenstein}, \citenamefont {Sanpera},\ and\ \citenamefont
  {Ahufinger}}]{lewenstein_ultracold_2012}%
  \BibitemOpen
  \bibfield  {author} {\bibinfo {author} {\bibfnamefont {M.}~\bibnamefont
  {Lewenstein}}, \bibinfo {author} {\bibfnamefont {A.}~\bibnamefont
  {Sanpera}},\ and\ \bibinfo {author} {\bibfnamefont {V.}~\bibnamefont
  {Ahufinger}},\ }\href@noop {} {\emph {\bibinfo {title} {Ultracold atoms in
  optical lattices: simulating quantum many-body systems}}},\ \bibinfo
  {edition} {1st}\ ed.\ (\bibinfo  {publisher} {Oxford University Press},\
  \bibinfo {address} {Oxford},\ \bibinfo {year} {2012})\BibitemShut {NoStop}%
\bibitem [{\citenamefont {Gross}\ and\ \citenamefont
  {Bloch}(2017)}]{gross_quantum_2017}%
  \BibitemOpen
  \bibfield  {author} {\bibinfo {author} {\bibfnamefont {C.}~\bibnamefont
  {Gross}}\ and\ \bibinfo {author} {\bibfnamefont {I.}~\bibnamefont {Bloch}},\
  }\href {https://doi.org/10.1126/science.aal3837} {\bibfield  {journal}
  {\bibinfo  {journal} {Science}\ }\textbf {\bibinfo {volume} {357}},\ \bibinfo
  {pages} {995} (\bibinfo {year} {2017})}\BibitemShut {NoStop}%
\bibitem [{\citenamefont {Jaksch}\ and\ \citenamefont
  {Zoller}(2005)}]{jaksch_cold_2005}%
  \BibitemOpen
  \bibfield  {author} {\bibinfo {author} {\bibfnamefont {D.}~\bibnamefont
  {Jaksch}}\ and\ \bibinfo {author} {\bibfnamefont {P.}~\bibnamefont
  {Zoller}},\ }\href {https://doi.org/10.1016/j.aop.2004.09.010} {\bibfield
  {journal} {\bibinfo  {journal} {Annals of Physics}\ }\textbf {\bibinfo
  {volume} {315}},\ \bibinfo {pages} {52} (\bibinfo {year} {2005})}\BibitemShut
  {NoStop}%
\bibitem [{\citenamefont {Chin}\ \emph {et~al.}(2010)\citenamefont {Chin},
  \citenamefont {Grimm}, \citenamefont {Julienne},\ and\ \citenamefont
  {Tiesinga}}]{chin_feshbach_2010}%
  \BibitemOpen
  \bibfield  {author} {\bibinfo {author} {\bibfnamefont {C.}~\bibnamefont
  {Chin}}, \bibinfo {author} {\bibfnamefont {R.}~\bibnamefont {Grimm}},
  \bibinfo {author} {\bibfnamefont {P.}~\bibnamefont {Julienne}},\ and\
  \bibinfo {author} {\bibfnamefont {E.}~\bibnamefont {Tiesinga}},\ }\href
  {https://doi.org/10.1103/revmodphys.82.1225} {\bibfield  {journal} {\bibinfo
  {journal} {Reviews of Modern Physics}\ }\textbf {\bibinfo {volume} {82}},\
  \bibinfo {pages} {1225} (\bibinfo {year} {2010})}\BibitemShut {NoStop}%
\bibitem [{\citenamefont {Morera}\ \emph {et~al.}(2020)\citenamefont {Morera},
  \citenamefont {Astrakharchik}, \citenamefont {Polls},\ and\ \citenamefont
  {Juliá-Díaz}}]{morera_quantum_2020}%
  \BibitemOpen
  \bibfield  {author} {\bibinfo {author} {\bibfnamefont {I.}~\bibnamefont
  {Morera}}, \bibinfo {author} {\bibfnamefont {G.~E.}\ \bibnamefont
  {Astrakharchik}}, \bibinfo {author} {\bibfnamefont {A.}~\bibnamefont
  {Polls}},\ and\ \bibinfo {author} {\bibfnamefont {B.}~\bibnamefont
  {Juliá-Díaz}},\ }\href {https://doi.org/10.1103/PhysRevResearch.2.022008}
  {\bibfield  {journal} {\bibinfo  {journal} {Physical Review Research}\
  }\textbf {\bibinfo {volume} {2}},\ \bibinfo {pages} {022008} (\bibinfo {year}
  {2020})}\BibitemShut {NoStop}%
\bibitem [{\citenamefont {Morera}\ \emph {et~al.}(2021)\citenamefont {Morera},
  \citenamefont {Astrakharchik}, \citenamefont {Polls},\ and\ \citenamefont
  {Juliá-Díaz}}]{morera_universal_2021}%
  \BibitemOpen
  \bibfield  {author} {\bibinfo {author} {\bibfnamefont {I.}~\bibnamefont
  {Morera}}, \bibinfo {author} {\bibfnamefont {G.~E.}\ \bibnamefont
  {Astrakharchik}}, \bibinfo {author} {\bibfnamefont {A.}~\bibnamefont
  {Polls}},\ and\ \bibinfo {author} {\bibfnamefont {B.}~\bibnamefont
  {Juliá-Díaz}},\ }\href {https://doi.org/10.1103/PhysRevLett.126.023001}
  {\bibfield  {journal} {\bibinfo  {journal} {Physical Review Letters}\
  }\textbf {\bibinfo {volume} {126}},\ \bibinfo {pages} {023001} (\bibinfo
  {year} {2021})}\BibitemShut {NoStop}%
\bibitem [{\citenamefont {Machida}\ \emph {et~al.}(2022)\citenamefont
  {Machida}, \citenamefont {Danshita}, \citenamefont {Yamamoto},\ and\
  \citenamefont {Kasamatsu}}]{machida_quantum_2022}%
  \BibitemOpen
  \bibfield  {author} {\bibinfo {author} {\bibfnamefont {Y.}~\bibnamefont
  {Machida}}, \bibinfo {author} {\bibfnamefont {I.}~\bibnamefont {Danshita}},
  \bibinfo {author} {\bibfnamefont {D.}~\bibnamefont {Yamamoto}},\ and\
  \bibinfo {author} {\bibfnamefont {K.}~\bibnamefont {Kasamatsu}},\ }\href
  {https://doi.org/10.1103/PhysRevA.105.L031301} {\bibfield  {journal}
  {\bibinfo  {journal} {Physical Review A}\ }\textbf {\bibinfo {volume}
  {105}},\ \bibinfo {pages} {L031301} (\bibinfo {year} {2022})}\BibitemShut
  {NoStop}%
\bibitem [{\citenamefont {Boudjemâa}\ and\ \citenamefont
  {Elhadj}(2023)}]{boudjemaa_discrete_2023}%
  \BibitemOpen
  \bibfield  {author} {\bibinfo {author} {\bibfnamefont {A.}~\bibnamefont
  {Boudjemâa}}\ and\ \bibinfo {author} {\bibfnamefont {K.~M.}\ \bibnamefont
  {Elhadj}},\ }\href {https://doi.org/10.1016/j.chaos.2023.114133} {\bibfield
  {journal} {\bibinfo  {journal} {Chaos, Solitons \& Fractals}\ }\textbf
  {\bibinfo {volume} {176}},\ \bibinfo {pages} {114133} (\bibinfo {year}
  {2023})}\BibitemShut {NoStop}%
\bibitem [{\citenamefont {Vallès-Muns}\ \emph {et~al.}(2024)\citenamefont
  {Vallès-Muns}, \citenamefont {Morera}, \citenamefont {Astrakharchik},\ and\
  \citenamefont {Juliá-Díaz}}]{valles-muns_quantum_2024}%
  \BibitemOpen
  \bibfield  {author} {\bibinfo {author} {\bibfnamefont {J.}~\bibnamefont
  {Vallès-Muns}}, \bibinfo {author} {\bibfnamefont {I.}~\bibnamefont
  {Morera}}, \bibinfo {author} {\bibfnamefont {G.~E.}\ \bibnamefont
  {Astrakharchik}},\ and\ \bibinfo {author} {\bibfnamefont {B.}~\bibnamefont
  {Juliá-Díaz}},\ }\href {https://doi.org/10.21468/SciPostPhys.16.3.074}
  {\bibfield  {journal} {\bibinfo  {journal} {SciPost Physics}\ }\textbf
  {\bibinfo {volume} {16}},\ \bibinfo {pages} {074} (\bibinfo {year}
  {2024})}\BibitemShut {NoStop}%
\bibitem [{\citenamefont {Rapp}\ \emph {et~al.}(2007)\citenamefont {Rapp},
  \citenamefont {Zarand}, \citenamefont {Honerkamp},\ and\ \citenamefont
  {Hofstetter}}]{rapp_color_2007}%
  \BibitemOpen
  \bibfield  {author} {\bibinfo {author} {\bibfnamefont {A.}~\bibnamefont
  {Rapp}}, \bibinfo {author} {\bibfnamefont {G.}~\bibnamefont {Zarand}},
  \bibinfo {author} {\bibfnamefont {C.}~\bibnamefont {Honerkamp}},\ and\
  \bibinfo {author} {\bibfnamefont {W.}~\bibnamefont {Hofstetter}},\ }\href
  {https://doi.org/10.1103/PhysRevLett.98.160405} {\bibfield  {journal}
  {\bibinfo  {journal} {Physical Review Letters}\ }\textbf {\bibinfo {volume}
  {98}},\ \bibinfo {pages} {160405} (\bibinfo {year} {2007})}\BibitemShut
  {NoStop}%
\bibitem [{\citenamefont {Rapp}\ \emph {et~al.}(2008)\citenamefont {Rapp},
  \citenamefont {Hofstetter},\ and\ \citenamefont
  {Zarand}}]{rapp_trionic_2008}%
  \BibitemOpen
  \bibfield  {author} {\bibinfo {author} {\bibfnamefont {A.}~\bibnamefont
  {Rapp}}, \bibinfo {author} {\bibfnamefont {W.}~\bibnamefont {Hofstetter}},\
  and\ \bibinfo {author} {\bibfnamefont {G.}~\bibnamefont {Zarand}},\ }\href
  {https://doi.org/10.1103/PhysRevB.77.144520} {\bibfield  {journal} {\bibinfo
  {journal} {Physical Review B}\ }\textbf {\bibinfo {volume} {77}},\ \bibinfo
  {pages} {144520} (\bibinfo {year} {2008})}\BibitemShut {NoStop}%
\bibitem [{\citenamefont {Klingschat}\ and\ \citenamefont
  {Honerkamp}(2010)}]{klingschat_exact_2010}%
  \BibitemOpen
  \bibfield  {author} {\bibinfo {author} {\bibfnamefont {G.}~\bibnamefont
  {Klingschat}}\ and\ \bibinfo {author} {\bibfnamefont {C.}~\bibnamefont
  {Honerkamp}},\ }\href {https://doi.org/10.1103/PhysRevB.82.094521} {\bibfield
   {journal} {\bibinfo  {journal} {Physical Review B}\ }\textbf {\bibinfo
  {volume} {82}},\ \bibinfo {pages} {094521} (\bibinfo {year}
  {2010})}\BibitemShut {NoStop}%
\bibitem [{\citenamefont {Backes}\ \emph {et~al.}(2012)\citenamefont {Backes},
  \citenamefont {Titvinidze}, \citenamefont {Privitera},\ and\ \citenamefont
  {Hofstetter}}]{backes_monte_2012}%
  \BibitemOpen
  \bibfield  {author} {\bibinfo {author} {\bibfnamefont {S.}~\bibnamefont
  {Backes}}, \bibinfo {author} {\bibfnamefont {I.}~\bibnamefont {Titvinidze}},
  \bibinfo {author} {\bibfnamefont {A.}~\bibnamefont {Privitera}},\ and\
  \bibinfo {author} {\bibfnamefont {W.}~\bibnamefont {Hofstetter}},\ }\href
  {https://doi.org/10.1103/PhysRevA.86.013633} {\bibfield  {journal} {\bibinfo
  {journal} {Physical Review A}\ }\textbf {\bibinfo {volume} {86}},\ \bibinfo
  {pages} {013633} (\bibinfo {year} {2012})}\BibitemShut {NoStop}%
\bibitem [{\citenamefont {Pohlmann}\ \emph {et~al.}(2013)\citenamefont
  {Pohlmann}, \citenamefont {Privitera}, \citenamefont {Titvinidze},\ and\
  \citenamefont {Hofstetter}}]{pohlmann_trion_2013}%
  \BibitemOpen
  \bibfield  {author} {\bibinfo {author} {\bibfnamefont {J.}~\bibnamefont
  {Pohlmann}}, \bibinfo {author} {\bibfnamefont {A.}~\bibnamefont {Privitera}},
  \bibinfo {author} {\bibfnamefont {I.}~\bibnamefont {Titvinidze}},\ and\
  \bibinfo {author} {\bibfnamefont {W.}~\bibnamefont {Hofstetter}},\ }\href
  {https://doi.org/10.1103/PhysRevA.87.023617} {\bibfield  {journal} {\bibinfo
  {journal} {Physical Review A}\ }\textbf {\bibinfo {volume} {87}},\ \bibinfo
  {pages} {023617} (\bibinfo {year} {2013})}\BibitemShut {NoStop}%
\bibitem [{\citenamefont {Xu}\ \emph {et~al.}(2023)\citenamefont {Xu},
  \citenamefont {Li}, \citenamefont {Zhou}, \citenamefont {Wang}, \citenamefont
  {Wang}, \citenamefont {Wu},\ and\ \citenamefont {Wang}}]{xu_trion_2023}%
  \BibitemOpen
  \bibfield  {author} {\bibinfo {author} {\bibfnamefont {H.}~\bibnamefont
  {Xu}}, \bibinfo {author} {\bibfnamefont {X.}~\bibnamefont {Li}}, \bibinfo
  {author} {\bibfnamefont {Z.}~\bibnamefont {Zhou}}, \bibinfo {author}
  {\bibfnamefont {X.}~\bibnamefont {Wang}}, \bibinfo {author} {\bibfnamefont
  {L.}~\bibnamefont {Wang}}, \bibinfo {author} {\bibfnamefont {C.}~\bibnamefont
  {Wu}},\ and\ \bibinfo {author} {\bibfnamefont {Y.}~\bibnamefont {Wang}},\
  }\href {https://doi.org/10.1103/PhysRevResearch.5.023180} {\bibfield
  {journal} {\bibinfo  {journal} {Physical Review Research}\ }\textbf {\bibinfo
  {volume} {5}},\ \bibinfo {pages} {023180} (\bibinfo {year}
  {2023})}\BibitemShut {NoStop}%
\bibitem [{\citenamefont {Zhang}\ and\ \citenamefont
  {Dong}(2010)}]{zhang_exact_2010}%
  \BibitemOpen
  \bibfield  {author} {\bibinfo {author} {\bibfnamefont {J.~M.}\ \bibnamefont
  {Zhang}}\ and\ \bibinfo {author} {\bibfnamefont {R.~X.}\ \bibnamefont
  {Dong}},\ }\href {https://doi.org/10.1088/0143-0807/31/3/016} {\bibfield
  {journal} {\bibinfo  {journal} {European Journal of Physics}\ }\textbf
  {\bibinfo {volume} {31}},\ \bibinfo {pages} {591} (\bibinfo {year}
  {2010})}\BibitemShut {NoStop}%
\bibitem [{\citenamefont {Raventós}\ \emph {et~al.}(2017)\citenamefont
  {Raventós}, \citenamefont {Graß}, \citenamefont {Lewenstein},\ and\
  \citenamefont {Juliá-Díaz}}]{raventos_cold_2017}%
  \BibitemOpen
  \bibfield  {author} {\bibinfo {author} {\bibfnamefont {D.}~\bibnamefont
  {Raventós}}, \bibinfo {author} {\bibfnamefont {T.}~\bibnamefont {Graß}},
  \bibinfo {author} {\bibfnamefont {M.}~\bibnamefont {Lewenstein}},\ and\
  \bibinfo {author} {\bibfnamefont {B.}~\bibnamefont {Juliá-Díaz}},\ }\href
  {https://doi.org/10.1088/1361-6455/aa68b1} {\bibfield  {journal} {\bibinfo
  {journal} {Journal of Physics B: Atomic, Molecular and Optical Physics}\
  }\textbf {\bibinfo {volume} {50}},\ \bibinfo {pages} {113001} (\bibinfo
  {year} {2017})}\BibitemShut {NoStop}%
\bibitem [{\citenamefont {Jaksch}\ \emph {et~al.}(1998)\citenamefont {Jaksch},
  \citenamefont {Bruder}, \citenamefont {Cirac}, \citenamefont {Gardiner},\
  and\ \citenamefont {Zoller}}]{jaksch_cold_1998}%
  \BibitemOpen
  \bibfield  {author} {\bibinfo {author} {\bibfnamefont {D.}~\bibnamefont
  {Jaksch}}, \bibinfo {author} {\bibfnamefont {C.}~\bibnamefont {Bruder}},
  \bibinfo {author} {\bibfnamefont {J.~I.}\ \bibnamefont {Cirac}}, \bibinfo
  {author} {\bibfnamefont {C.~W.}\ \bibnamefont {Gardiner}},\ and\ \bibinfo
  {author} {\bibfnamefont {P.}~\bibnamefont {Zoller}},\ }\href
  {https://doi.org/10.1103/PhysRevLett.81.3108} {\bibfield  {journal} {\bibinfo
   {journal} {Physical Review Letters}\ }\textbf {\bibinfo {volume} {81}},\
  \bibinfo {pages} {3108} (\bibinfo {year} {1998})}\BibitemShut {NoStop}%
\bibitem [{\citenamefont {Amelio}\ \emph {et~al.}(2024)\citenamefont {Amelio},
  \citenamefont {Mazza},\ and\ \citenamefont {Goldman}}]{amelio_polaron_2024}%
  \BibitemOpen
  \bibfield  {author} {\bibinfo {author} {\bibfnamefont {I.}~\bibnamefont
  {Amelio}}, \bibinfo {author} {\bibfnamefont {G.}~\bibnamefont {Mazza}},\ and\
  \bibinfo {author} {\bibfnamefont {N.}~\bibnamefont {Goldman}},\ }\href
  {https://doi.org/10.1103/PhysRevB.110.235302} {\bibfield  {journal} {\bibinfo
   {journal} {Physical Review B}\ }\textbf {\bibinfo {volume} {110}},\ \bibinfo
  {pages} {235302} (\bibinfo {year} {2024})}\BibitemShut {NoStop}%
\bibitem [{\citenamefont {Lehoucq}\ \emph {et~al.}(1998)\citenamefont
  {Lehoucq}, \citenamefont {Sorensen},\ and\ \citenamefont
  {Yang}}]{lehoucq_arpack_1998}%
  \BibitemOpen
  \bibfield  {author} {\bibinfo {author} {\bibfnamefont {R.~B.}\ \bibnamefont
  {Lehoucq}}, \bibinfo {author} {\bibfnamefont {D.~C.}\ \bibnamefont
  {Sorensen}},\ and\ \bibinfo {author} {\bibfnamefont {C.}~\bibnamefont
  {Yang}},\ }\href@noop {} {\emph {\bibinfo {title} {Arpack {Users}’ {Guide}:
  {Solution} of {Large} {Scale} {Eigenvalue} {Problems} with {Implicitly}
  {Restarted} {Arnoldi} {Methods}}}}\ (\bibinfo  {publisher} {SIAM},\ \bibinfo
  {address} {Philadelphia},\ \bibinfo {year} {1998})\BibitemShut {NoStop}%
\bibitem [{\citenamefont {Hadzibabic}\ \emph {et~al.}(2008)\citenamefont
  {Hadzibabic}, \citenamefont {Krüger}, \citenamefont {Cheneau}, \citenamefont
  {Rath},\ and\ \citenamefont {Dalibard}}]{hadzibabic_trapped_2008}%
  \BibitemOpen
  \bibfield  {author} {\bibinfo {author} {\bibfnamefont {Z.}~\bibnamefont
  {Hadzibabic}}, \bibinfo {author} {\bibfnamefont {P.}~\bibnamefont {Krüger}},
  \bibinfo {author} {\bibfnamefont {M.}~\bibnamefont {Cheneau}}, \bibinfo
  {author} {\bibfnamefont {S.~P.}\ \bibnamefont {Rath}},\ and\ \bibinfo
  {author} {\bibfnamefont {J.}~\bibnamefont {Dalibard}},\ }\href
  {https://doi.org/10.1088/1367-2630/10/4/045006} {\bibfield  {journal}
  {\bibinfo  {journal} {New Journal of Physics}\ }\textbf {\bibinfo {volume}
  {10}},\ \bibinfo {pages} {045006} (\bibinfo {year} {2008})}\BibitemShut
  {NoStop}%
\bibitem [{\citenamefont {Zhang}\ \emph {et~al.}(2012)\citenamefont {Zhang},
  \citenamefont {Hung}, \citenamefont {Tung},\ and\ \citenamefont
  {Chin}}]{zhang_observation_2012}%
  \BibitemOpen
  \bibfield  {author} {\bibinfo {author} {\bibfnamefont {X.}~\bibnamefont
  {Zhang}}, \bibinfo {author} {\bibfnamefont {C.-L.}\ \bibnamefont {Hung}},
  \bibinfo {author} {\bibfnamefont {S.-K.}\ \bibnamefont {Tung}},\ and\
  \bibinfo {author} {\bibfnamefont {C.}~\bibnamefont {Chin}},\ }\href
  {https://doi.org/10.1126/science.1217990} {\bibfield  {journal} {\bibinfo
  {journal} {Science}\ }\textbf {\bibinfo {volume} {335}},\ \bibinfo {pages}
  {1070} (\bibinfo {year} {2012})}\BibitemShut {NoStop}%
\bibitem [{\citenamefont {Cocchi}\ \emph {et~al.}(2016)\citenamefont {Cocchi},
  \citenamefont {Miller}, \citenamefont {Drewes}, \citenamefont {Koschorreck},
  \citenamefont {Pertot}, \citenamefont {Brennecke},\ and\ \citenamefont
  {Köhl}}]{cocchi_equation_2016}%
  \BibitemOpen
  \bibfield  {author} {\bibinfo {author} {\bibfnamefont {E.}~\bibnamefont
  {Cocchi}}, \bibinfo {author} {\bibfnamefont {L.~A.}\ \bibnamefont {Miller}},
  \bibinfo {author} {\bibfnamefont {J.~H.}\ \bibnamefont {Drewes}}, \bibinfo
  {author} {\bibfnamefont {M.}~\bibnamefont {Koschorreck}}, \bibinfo {author}
  {\bibfnamefont {D.}~\bibnamefont {Pertot}}, \bibinfo {author} {\bibfnamefont
  {F.}~\bibnamefont {Brennecke}},\ and\ \bibinfo {author} {\bibfnamefont
  {M.}~\bibnamefont {Köhl}},\ }\href
  {https://doi.org/10.1103/PhysRevLett.116.175301} {\bibfield  {journal}
  {\bibinfo  {journal} {Physical Review Letters}\ }\textbf {\bibinfo {volume}
  {116}},\ \bibinfo {pages} {175301} (\bibinfo {year} {2016})}\BibitemShut
  {NoStop}%
\bibitem [{\citenamefont {Blakie}\ and\ \citenamefont
  {Clark}(2004)}]{blakie_wannier_2004}%
  \BibitemOpen
  \bibfield  {author} {\bibinfo {author} {\bibfnamefont {P.~B.}\ \bibnamefont
  {Blakie}}\ and\ \bibinfo {author} {\bibfnamefont {C.~W.}\ \bibnamefont
  {Clark}},\ }\href {https://doi.org/10.1088/0953-4075/37/7/002} {\bibfield
  {journal} {\bibinfo  {journal} {Journal of Physics B: Atomic, Molecular and
  Optical Physics}\ }\textbf {\bibinfo {volume} {37}},\ \bibinfo {pages} {1391}
  (\bibinfo {year} {2004})}\BibitemShut {NoStop}%
\bibitem [{\citenamefont {Walters}\ \emph {et~al.}(2013)\citenamefont
  {Walters}, \citenamefont {Cotugno}, \citenamefont {Johnson}, \citenamefont
  {Clark},\ and\ \citenamefont {Jaksch}}]{walters_ab_2013}%
  \BibitemOpen
  \bibfield  {author} {\bibinfo {author} {\bibfnamefont {R.}~\bibnamefont
  {Walters}}, \bibinfo {author} {\bibfnamefont {G.}~\bibnamefont {Cotugno}},
  \bibinfo {author} {\bibfnamefont {T.~H.}\ \bibnamefont {Johnson}}, \bibinfo
  {author} {\bibfnamefont {S.~R.}\ \bibnamefont {Clark}},\ and\ \bibinfo
  {author} {\bibfnamefont {D.}~\bibnamefont {Jaksch}},\ }\href
  {https://doi.org/10.1103/PhysRevA.87.043613} {\bibfield  {journal} {\bibinfo
  {journal} {Physical Review A}\ }\textbf {\bibinfo {volume} {87}},\ \bibinfo
  {pages} {043613} (\bibinfo {year} {2013})}\BibitemShut {NoStop}%
\bibitem [{\citenamefont {Isaule}\ \emph {et~al.}(2024)\citenamefont {Isaule},
  \citenamefont {Rojo-Francàs},\ and\ \citenamefont
  {Juliá-Díaz}}]{isaule_bound_2024}%
  \BibitemOpen
  \bibfield  {author} {\bibinfo {author} {\bibfnamefont {F.}~\bibnamefont
  {Isaule}}, \bibinfo {author} {\bibfnamefont {A.}~\bibnamefont
  {Rojo-Francàs}},\ and\ \bibinfo {author} {\bibfnamefont {B.}~\bibnamefont
  {Juliá-Díaz}},\ }\href {https://doi.org/10.21468/SciPostPhysCore.7.3.049}
  {\bibfield  {journal} {\bibinfo  {journal} {SciPost Physics Core}\ }\textbf
  {\bibinfo {volume} {7}},\ \bibinfo {pages} {049} (\bibinfo {year}
  {2024})}\BibitemShut {NoStop}%
\bibitem [{\citenamefont {Amico}\ \emph {et~al.}(2008)\citenamefont {Amico},
  \citenamefont {Fazio}, \citenamefont {Osterloh},\ and\ \citenamefont
  {Vedral}}]{amico_entanglement_2008}%
  \BibitemOpen
  \bibfield  {author} {\bibinfo {author} {\bibfnamefont {L.}~\bibnamefont
  {Amico}}, \bibinfo {author} {\bibfnamefont {R.}~\bibnamefont {Fazio}},
  \bibinfo {author} {\bibfnamefont {A.}~\bibnamefont {Osterloh}},\ and\
  \bibinfo {author} {\bibfnamefont {V.}~\bibnamefont {Vedral}},\ }\href
  {https://doi.org/10.1103/RevModPhys.80.517} {\bibfield  {journal} {\bibinfo
  {journal} {Reviews of Modern Physics}\ }\textbf {\bibinfo {volume} {80}},\
  \bibinfo {pages} {517} (\bibinfo {year} {2008})}\BibitemShut {NoStop}%
\bibitem [{\citenamefont {Brand}\ \emph {et~al.}(2008)\citenamefont {Brand},
  \citenamefont {Flach}, \citenamefont {Fleurov}, \citenamefont {Schulman},\
  and\ \citenamefont {Tolkunov}}]{brand_localization_2008}%
  \BibitemOpen
  \bibfield  {author} {\bibinfo {author} {\bibfnamefont {J.}~\bibnamefont
  {Brand}}, \bibinfo {author} {\bibfnamefont {S.}~\bibnamefont {Flach}},
  \bibinfo {author} {\bibfnamefont {V.}~\bibnamefont {Fleurov}}, \bibinfo
  {author} {\bibfnamefont {L.~S.}\ \bibnamefont {Schulman}},\ and\ \bibinfo
  {author} {\bibfnamefont {D.}~\bibnamefont {Tolkunov}},\ }\href
  {https://doi.org/10.1209/0295-5075/83/40002} {\bibfield  {journal} {\bibinfo
  {journal} {EPL (Europhysics Letters)}\ }\textbf {\bibinfo {volume} {83}},\
  \bibinfo {pages} {40002} (\bibinfo {year} {2008})}\BibitemShut {NoStop}%
\bibitem [{\citenamefont {Mujal}\ \emph {et~al.}(2016)\citenamefont {Mujal},
  \citenamefont {Juliá-Díaz},\ and\ \citenamefont
  {Polls}}]{mujal_quantum_2016}%
  \BibitemOpen
  \bibfield  {author} {\bibinfo {author} {\bibfnamefont {P.}~\bibnamefont
  {Mujal}}, \bibinfo {author} {\bibfnamefont {B.}~\bibnamefont
  {Juliá-Díaz}},\ and\ \bibinfo {author} {\bibfnamefont {A.}~\bibnamefont
  {Polls}},\ }\href {https://doi.org/10.1103/PhysRevA.93.043619} {\bibfield
  {journal} {\bibinfo  {journal} {Physical Review A}\ }\textbf {\bibinfo
  {volume} {93}},\ \bibinfo {pages} {043619} (\bibinfo {year}
  {2016})}\BibitemShut {NoStop}%
\bibitem [{\citenamefont {Richaud}\ and\ \citenamefont
  {Penna}(2019)}]{richaud_pathway_2019}%
  \BibitemOpen
  \bibfield  {author} {\bibinfo {author} {\bibfnamefont {A.}~\bibnamefont
  {Richaud}}\ and\ \bibinfo {author} {\bibfnamefont {V.}~\bibnamefont
  {Penna}},\ }\href {https://doi.org/10.1103/PhysRevA.100.013609} {\bibfield
  {journal} {\bibinfo  {journal} {Physical Review A}\ }\textbf {\bibinfo
  {volume} {100}},\ \bibinfo {pages} {013609} (\bibinfo {year}
  {2019})}\BibitemShut {NoStop}%
\bibitem [{\citenamefont {Gubernatis}\ \emph {et~al.}(2016)\citenamefont
  {Gubernatis}, \citenamefont {Kawashima},\ and\ \citenamefont
  {Werner}}]{gubernatis_quantum_2016}%
  \BibitemOpen
  \bibfield  {author} {\bibinfo {author} {\bibfnamefont {J.}~\bibnamefont
  {Gubernatis}}, \bibinfo {author} {\bibfnamefont {N.}~\bibnamefont
  {Kawashima}},\ and\ \bibinfo {author} {\bibfnamefont {P.}~\bibnamefont
  {Werner}},\ }\href {https://doi.org/10.1017/CBO9780511902581} {\emph
  {\bibinfo {title} {Quantum {Monte} {Carlo} {Methods}: {Algorithms} for
  {Lattice} {Models}}}},\ \bibinfo {edition} {1st}\ ed.\ (\bibinfo  {publisher}
  {Cambridge University Press},\ \bibinfo {year} {2016})\BibitemShut {NoStop}%
\bibitem [{\citenamefont {Motoyama}\ \emph {et~al.}(2021)\citenamefont
  {Motoyama}, \citenamefont {Yoshimi}, \citenamefont {Masaki-Kato},
  \citenamefont {Kato},\ and\ \citenamefont {Kawashima}}]{motoyama_dsqss_2021}%
  \BibitemOpen
  \bibfield  {author} {\bibinfo {author} {\bibfnamefont {Y.}~\bibnamefont
  {Motoyama}}, \bibinfo {author} {\bibfnamefont {K.}~\bibnamefont {Yoshimi}},
  \bibinfo {author} {\bibfnamefont {A.}~\bibnamefont {Masaki-Kato}}, \bibinfo
  {author} {\bibfnamefont {T.}~\bibnamefont {Kato}},\ and\ \bibinfo {author}
  {\bibfnamefont {N.}~\bibnamefont {Kawashima}},\ }\href
  {https://doi.org/10.1016/j.cpc.2021.107944} {\bibfield  {journal} {\bibinfo
  {journal} {Computer Physics Communications}\ }\textbf {\bibinfo {volume}
  {264}},\ \bibinfo {pages} {107944} (\bibinfo {year} {2021})}\BibitemShut
  {NoStop}%
\bibitem [{\citenamefont {Caleffi}\ \emph {et~al.}(2020)\citenamefont
  {Caleffi}, \citenamefont {Capone}, \citenamefont {Menotti}, \citenamefont
  {Carusotto},\ and\ \citenamefont {Recati}}]{caleffi_quantum_2020}%
  \BibitemOpen
  \bibfield  {author} {\bibinfo {author} {\bibfnamefont {F.}~\bibnamefont
  {Caleffi}}, \bibinfo {author} {\bibfnamefont {M.}~\bibnamefont {Capone}},
  \bibinfo {author} {\bibfnamefont {C.}~\bibnamefont {Menotti}}, \bibinfo
  {author} {\bibfnamefont {I.}~\bibnamefont {Carusotto}},\ and\ \bibinfo
  {author} {\bibfnamefont {A.}~\bibnamefont {Recati}},\ }\href
  {https://doi.org/10.1103/PhysRevResearch.2.033276} {\bibfield  {journal}
  {\bibinfo  {journal} {Physical Review Research}\ }\textbf {\bibinfo {volume}
  {2}},\ \bibinfo {pages} {033276} (\bibinfo {year} {2020})}\BibitemShut
  {NoStop}%
\bibitem [{\citenamefont {Colussi}\ \emph {et~al.}(2022)\citenamefont
  {Colussi}, \citenamefont {Caleffi}, \citenamefont {Menotti},\ and\
  \citenamefont {Recati}}]{colussi_quantum_2022}%
  \BibitemOpen
  \bibfield  {author} {\bibinfo {author} {\bibfnamefont {V.}~\bibnamefont
  {Colussi}}, \bibinfo {author} {\bibfnamefont {F.}~\bibnamefont {Caleffi}},
  \bibinfo {author} {\bibfnamefont {C.}~\bibnamefont {Menotti}},\ and\ \bibinfo
  {author} {\bibfnamefont {A.}~\bibnamefont {Recati}},\ }\href
  {https://doi.org/10.21468/SciPostPhys.12.3.111} {\bibfield  {journal}
  {\bibinfo  {journal} {SciPost Physics}\ }\textbf {\bibinfo {volume} {12}},\
  \bibinfo {pages} {111} (\bibinfo {year} {2022})}\BibitemShut {NoStop}%
\bibitem [{Note1()}]{Note1}%
  \BibitemOpen
  \bibinfo {note} {Data for: Data for: Few-body bound states of bosonic
  mixtures in two-dimensional optical lattices (2025), Zenodo, \protect \href
  {http://doi.org/10.5281/zenodo.15775764}{10.5281/zenodo.15775764}.}\BibitemShut
  {Stop}%
\end{thebibliography}%

\end{document}